\documentclass[12pt,preprint]{aastex}
\usepackage{verbatim}

\shorttitle{Hierarchical Structure of MHD Turbulence}

\shortauthors{BURKHART ET AL.}
\begin{document}

\title{Hierarchical Structure of Magnetohydrodynamic Turbulence In Position-Position-Velocity Space}

\author{ Blakesley Burkhart\altaffilmark{1}, A. Lazarian\altaffilmark{1}, Alyssa Goodman\altaffilmark{2}, Erik Rosolowsky\altaffilmark{3}}
\altaffiltext{1}{Astronomy Department, University of Wisconsin, Madison, 475 N. 
Charter St., WI 53711, USA}
\altaffiltext{2}{ Harvard-Smithsonian Center for Astrophysics, 60 Garden Street, MS-78, Cambridge, MA 02138}
\altaffiltext{3}{University of British Columbia, Okanagan Campus, 3333 University Way, Kelowna BC V1V 1V7, Canada }

\begin{abstract}
Magnetohydrodynamic turbulence is able to create hierarchical structures in the interstellar medium that are correlated on a wide range of scales via
the energy cascade.  We use hierarchical tree diagrams known as dendrograms to characterize structures
in synthetic Position-Position-Velocity (PPV) emission cubes of isothermal magnetohydrodynamic turbulence. 
We show that the structures and degree of hierarchy observed in PPV space are related to the presence of
self-gravity and the global sonic and Alfv\'enic Mach numbers.
Simulations with higher Alfv\'enic Mach number, self-gravity and supersonic flows display enhanced hierarchical structure.  
We observe a strong dependency on the sonic and Alfv\'enic Mach numbers and self-gravity when we apply the statistical moments (i.e. mean, variance, skewness, kurtosis)
to the leaf and node distribution of the dendrogram. Simulations with self-gravity, larger magnetic field and higher sonic Mach number have dendrogram distributions with higher statistical moments.
Application of the dendrogram to 3D density cubes, also known as Position-Position-Position cubes (PPP), reveals that
the dominant emission contours in PPP and PPV are related for supersonic gas but not for subsonic. 
We also explore the effects of smoothing, thermal broadening and velocity resolution on the dendrograms in order to make our study more applicable to observational data.
These results all point to hierarchical tree diagrams as being a promising additional tool for studying
ISM turbulence and star forming regions for obtaining information on the degree of self-gravity, the Mach numbers and the complicated relationship between PPV and PPP data.
\end{abstract}
 \keywords{ISM: structure --- MHD --- turbulence}

\section{Introduction}
\label{intro}
 The current understanding of the interstellar medium (ISM) is that it is a multi-phase environment, consisting of gas and dust, which is both magnetized and highly turbulent (Ferriere 2001; McKee \& Ostriker 2007).
 In particular,  magnetohydrodynamic (MHD) turbulence is essential
to many astrophysical phenomena such as star formation, cosmic ray dispersion, and many transport processes
(see Elmegreen \& Scalo 2004; Ballesteros-Paredes et al. 2007 and references therein).
Additionally, turbulence has the unique ability to transfer energy over scales ranging from kiloparsecs down to the proton gyroradius.
This is critical for the ISM, as it explains how energy is distributed from large to small spatial scales in the Galaxy.

Observationally, several techniques exist to study MHD turbulence in different ISM phases.  Many of these techniques focus on  
emission measure fluctuations or  rotation measure fluctuations (i.e gradients of linear polarization maps) for the warm ionized media (Armstrong et al.1995; Chepurnov \& Lazarian 2010; Gaensler et al. 2011; Burkhart, Lazarian \& Gaensler 2012) as well as
spectroscopic data and column density maps for the neutral warm and cold media  (Spangler \& Gwinn 1990; Padoan et al. 2003), 
For studies of turbulence,  spectroscopic data has a clear advantage in that it contains information about the 
turbulent velocity field as well as the density fluctuations. However, density and velocity are entangled in PPV space, making the interpretation of this
type of data difficult.  For the separation of the density and velocity
fluctuations, special techniques such as the Velocity Coordinate Spectrum (VCS) and the Velocity Channel Analysis (VCA)
have been developed (Lazarian \& Pogosyan 2000, 2004, 2006, 2008).

Most of the efforts to relate observations and simulations of magnetized turbulence are based on obtaining the spectral
index (i.e. the log-log slope of the power spectrum) of either the density and/or velocity (Lazarian \& Esquivel 2003; Esquivel \& Lazarian 2005; Ossenkopf et al. 2006). 
However, the power spectrum alone does not provide a full description of turbulence, as it only contains the
Fourier amplitudes and neglects information on phases.  This fact,
combined with complexity of  astrophysical turbulence, with
multiple injection scales occurring in a multiphase medium, suggests researchers need additional ways of analyzing 
observational and numerical data in the context of turbulence.  In particular, these technique studies are currently focused into
two categories:

\begin{itemize}
 \item Development: Test and develop techniques that will complement and build off the theoretical and practical picture
 of a turbulent ISM that the power spectrum presents.
\item Synergy: Use several techniques simultaneously to obtain an accurate picture of the parameters of turbulence in the observations.
\end{itemize}

In regards to the first point, in the last decade there has been substantial progress in the development of techniques to study turbulence. Techniques for the study of turbulence can be tested empirically,
using parameter studies of numerical simulations, 
or with the aid of analytical predictions (as was done in the case of the VCA).  In the former, the parameters to be varied (see Burkhart \& Lazarian
2011) include the Reynolds number, sonic and Alfv\'enic Mach number, injection scale, equation of state, and, for studies of molecular clouds,
should include radiative transfer and self-gravity (see Ossenkopf 2002; Padoan et al. 2003; Goodman et al. 2009). 
Some recently developed techniques include the application of
probability distribution functions (PDFs), wavelets, spectral correlation function (SCF),\footnote{The similarities between VCA and SCF are discussed in Lazarian (2009).}
delta-variance, the principal
component analysis,  higher order moments, Genus, Tsallis statistics, spectrum and bispectrum (Gill \& Henriksen
1990; Stutzki et al. 1998; Rosolowsky et al. 1999; Brunt \& Heyer 2002; Kowal, Lazarian \& Beresnyak 2007;  Chepurnov et al. 2008; 
Burkhart et al. 2009; Esquivel \& Lazarian 2010; Tofflemire et al. 2011). 
Additionally, these techniques are being tested and  applied to
different  wavelengths and types of data.  For example, the  PDFs and their mathematical descriptors have been applied to the \textit{observations}
in the context of turbulence in numerous works using linear 
polarization data (see Gaensler et al. 2011; Burkhart, Lazarian, \& Gaensler 2012), HI column density of the SMC (Burkhart et al. 2010), molecular/dust extinction maps
(Goodman, Pineda, \& Schnee 2009; Brunt 2010; Kainulainen et al. 2011) and 
emission measure and volume averaged density in diffuse ionized gas (Hill et al. 2008; Berkhuijsen \& Fletcher 2008).

The latter point in regard to the synergetic use of tools for ISM turbulence is only recently being attempted as many techniques are still in developmental stages. 
However, this approach was used in Burkhart et al. (2010), which applied spectrum, bispectrum and higher 
order moments to HI column density of the SMC.  The consistency of results obtained with a variety
of statistics, compared with more traditional observational methods, made this study of turbulence in the SMC 
a promising first step. 
 
The current paper falls under the category of ``technique development.'' 
In particular, we investigate the utility of dendrograms in studying the hierarchical structure of ISM clouds.    
It has long been known that turbulence is able to create hierarchical structures in the ISM (Scalo 1985, 1990; Vazquez-Semadeni 1994; Stutzki 1998), however many questions remain, such as what type of turbulence
is behind the creation of this hierarchy and what is the role of self-gravity and magnetic fields?
Hierarchical structure in relation to this questions is particularly important for the star formation problem
(Larson 1981; Elmegreen \& Elmegreen 1983; Feitzinger \& Galinski 1987; Elmegreen 1999; Elmegreen 2011).

\begin{figure}[tbh]
\centering
\includegraphics[scale=1]{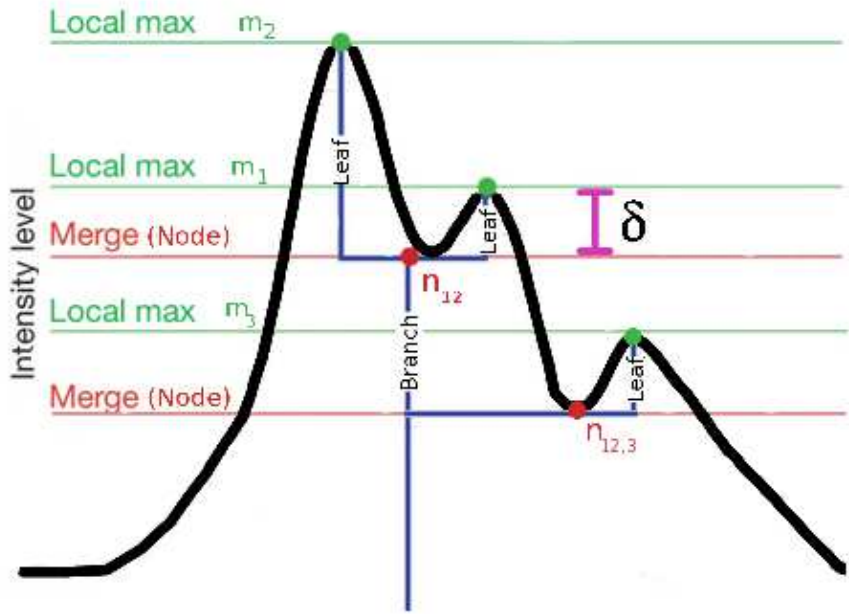}
\caption{The dendrogram for a hypothetical 1D emission profile showing three local maximum (leaves) and  merger points (nodes). The Dendrogram is shown
in blue and can be altered by changing the threshold level $\delta$ to higher or lower values.  In this example, increasing the value
of $\delta$ will merge the smallest leaf into the larger structure. The local maximum (green dots) and merger points (i.e. nodes, red dot)
are the values used to create the distribution $\xi$, discussed further in Section \S~\ref{results}. }
\label{fig:alyssa}
\end{figure}

Early attempts to characterize ISM hierarchy utilized tree diagrams as a mechanism for reducing the data
to hierarchical ``skeleton images'' (see Houlahan \& Scalo 1992). 
More recently, dendrograms have been used on ISM data  in order to characterize self-gravitating structures in star forming molecular 
clouds (Rosolowsky et al. 2008 and Goodman et al. 2009). 
A dendrogram (from the Greek dendron ``tree'',- gramma ``drawing'') is a hierarchical tree diagram 
that has been used extensively in other fields, particularly in computational biology, and occasionally in galaxy evolution
(see Sawlaw \& Haque-Copilah 1998 and Podani, Engloner, \& Major 2009, for examples).
Rosolowsky et al. (2008) and Goodman et al. (2009)  used the dendrogram on  spectral line data of 
L1448 to estimate key physical properties associated with isosurfaces of local emission maxima such 
as radius, velocity dispersion, and luminosity. These works provided a new and promising
way of characterizing self-gravitating structures and properties of molecular clouds through the application
of dendrogram to $^{13}$CO(J=1-0) PPV data.

In the current paper we apply the dendrogram to synthetic observations (specifically PPV cubes) of isothermal MHD turbulence in order to investigate the physical mechanisms
behind the gas hierarchy. Additionally, we are interested in the nature of the structures that are found in PPV data and how these structures
are related to both the physics of the gas and the underlying density and velocity fluctuations generated by turbulence. Simulations provide an excellent testing ground
for this problem, as one can identify which features in PPV space are density features and which  are caused by velocity crowding.  
Furthermore, one can answer the question: under what conditions do the features  in PPV relate back to the 3D density or PPP cube?

In order to address these questions we perform a parameter study using the dendrogram. We focus on how changing the global parameters of the turbulence, 
such as the sonic Mach number, Alfv\'enic Mach number, affects the amount of hierarchy observed, the relationship between the density and velocity
structures in PPV,  and the number and statistical distribution of dominant emission structures.

The paper is organized as follows.  In \S~\ref{dendoalg} we describe the dendrogram algorithm,
in \S~\ref{data} we discuss the simulations and provide a description of the MHD models.
We investigate the physical mechanisms that create hierarchical structure 
in the dendrogram tree and  characterize the tree diagrams via  statistical moments 
 in \S~\ref{results}. In  \S~\ref{ppp} we compare the dendrograms
of PPP and PPV.  In \S~\ref{sec:app} we discuss applications and investigate issues of resolution.
Finally, in  \S~\ref{disc} we discuss our results followed by the conclusions in \S~\ref{con}. 

\section{Dendrogram Algorithm}
\label{dendoalg}

The dendrogram is a tree diagram that can be used in 1D, 2D or 3D spaces to characterize how and where local
maxima merge as a function of a threshold parameter.  Although the current paper uses the dendrogram in 3D PPV space to characterize the merger of local maxima of emission,
 it is more intuitive to understand
the 1D and 2D applications. A 1D example of the dendrogram algorithm for an emission profile is shown in Figure \ref{fig:alyssa}.
In this case, the threshold value is called $\delta$, and is the minimum amplitude above a merger point that a local
maximum must have before it is considered distinct. That is, if a merger point (or node) is given by $n$ and a local maximum is given by $m$ then 
in order for a given local max $m_{1}$ to be considered significant, $m_{1}-n_{1,2} > \delta$.  If $m_{1}-n_{1,2} \le \delta$
then $m_{1}$ would merge into $m_{2}$ and no longer be considered distinct.

For 2D data, a common analogy (see Houlahan \& Scalo 1992; Rosolowsky et al. 2008)
is to think of the dendrogram technique as a descriptor of an underwater mountain chain. As the water level is lowered,
first one would see the peaks of the mountain, then mountain valleys (saddle points) and as more water is drained, the peaks may
merge together into larger objects.  The dendrogram stores information about the peaks and merger levels of the mountain
chain.

\begin{figure}[tlh]
\centering
\includegraphics[scale=.73]{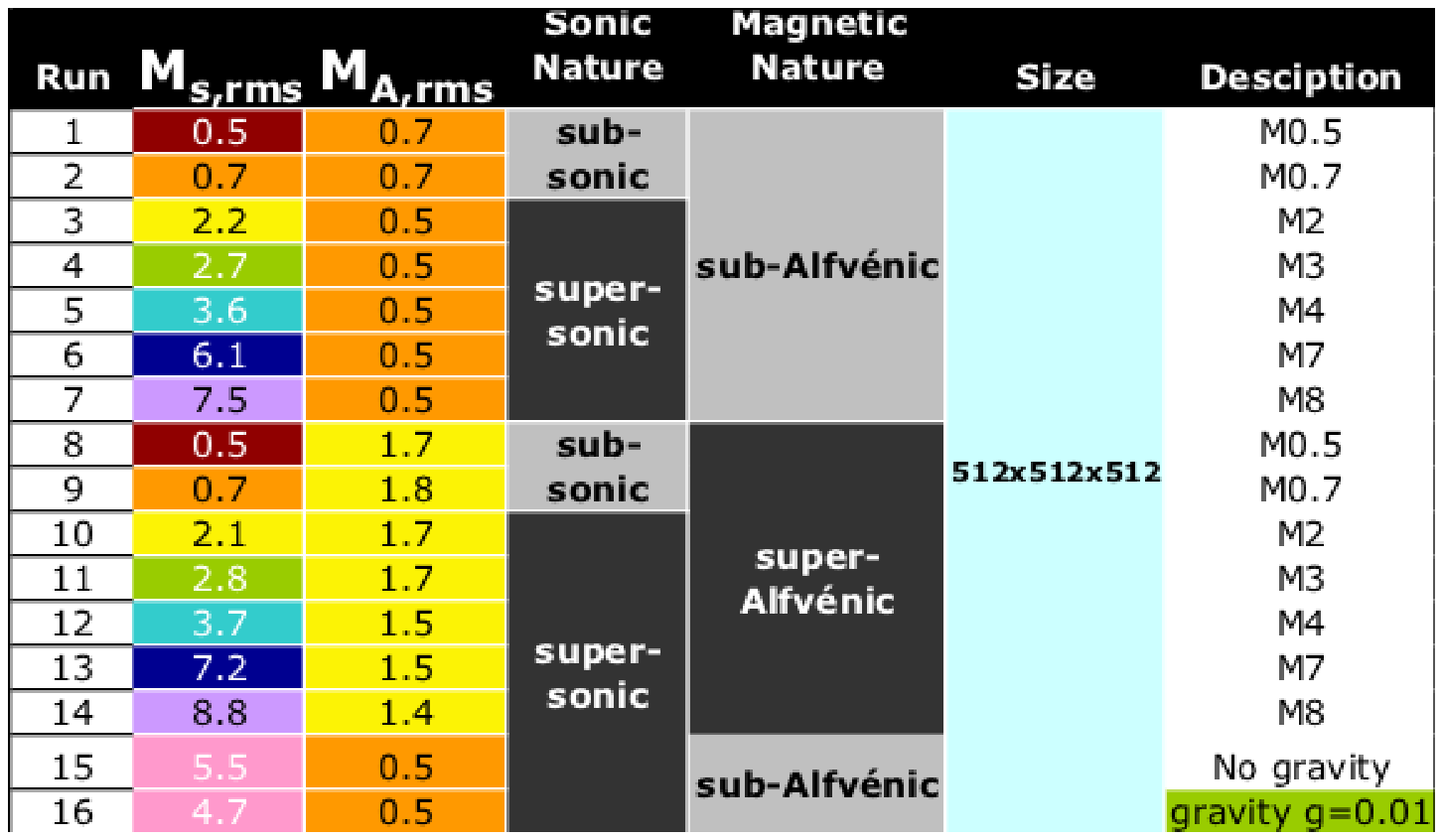}
\caption{List of the simulations and their properties. We define
the subsonic regimes as anything less than ${\cal M}_s$=1  and the supersonic regime as
${\cal M}_s  > 1$.   Two Alfv\'enic regimes exist for each sonic Mach number: super-Alfv\'enic and sub-Alfv\'enic. Same color along the ${\cal M}_s$ column
indicates that the same initial sound speed was used.  Same color along the ${\cal M}_A$ column indicates that the same initial mean field strength was used.
We group the description of the $512^3$ simulations based on similar values of the initial sound speed.  For example, Run 1 and Run 8 have the same initial sound speed 
and are both described as M0.5.}
\label{descrp}
\end{figure}

For our purposes, we examine the dendrogram in 3D PPV space (see Rosolowsky et al. 2008; Goodman et al. 2009 for more information on the 
dendrogram algorithm applied in PPV).  In the 3D case,
it is useful to think of each point in the dendrogram as representing a 3D contour (isosurface) in the data
cube at a given level. 

Our implementation of the dendrogram  is similar to many other statistics that employ a user defined
threshold value in order to classify structure. By varying the threshold
parameter for  the definition of  ``local maximum''  (which we call $\delta$, see Figure \ref{fig:alyssa}),  different dendrogram tree diagrams and  distributions of 
local maximum and merger points  are 
created.  An example of another statistic
that utilizes a density/emission threshold value is the Genus statistic, 
which has proven useful for studying ISM topology
(Lazarian, Pogosyan \& Esquivel 2002; Lazarian 2004; Kim \& Park 2007; Kowal, Lazarian \& Beresnyak 2007;  Chepurnov et al. 2008). 
For the Genus technique, the variation of the threshold value is a critical point in understanding the 
topology of the data in question.

\begin{figure*}[th]
\centering
\includegraphics[scale=.9]{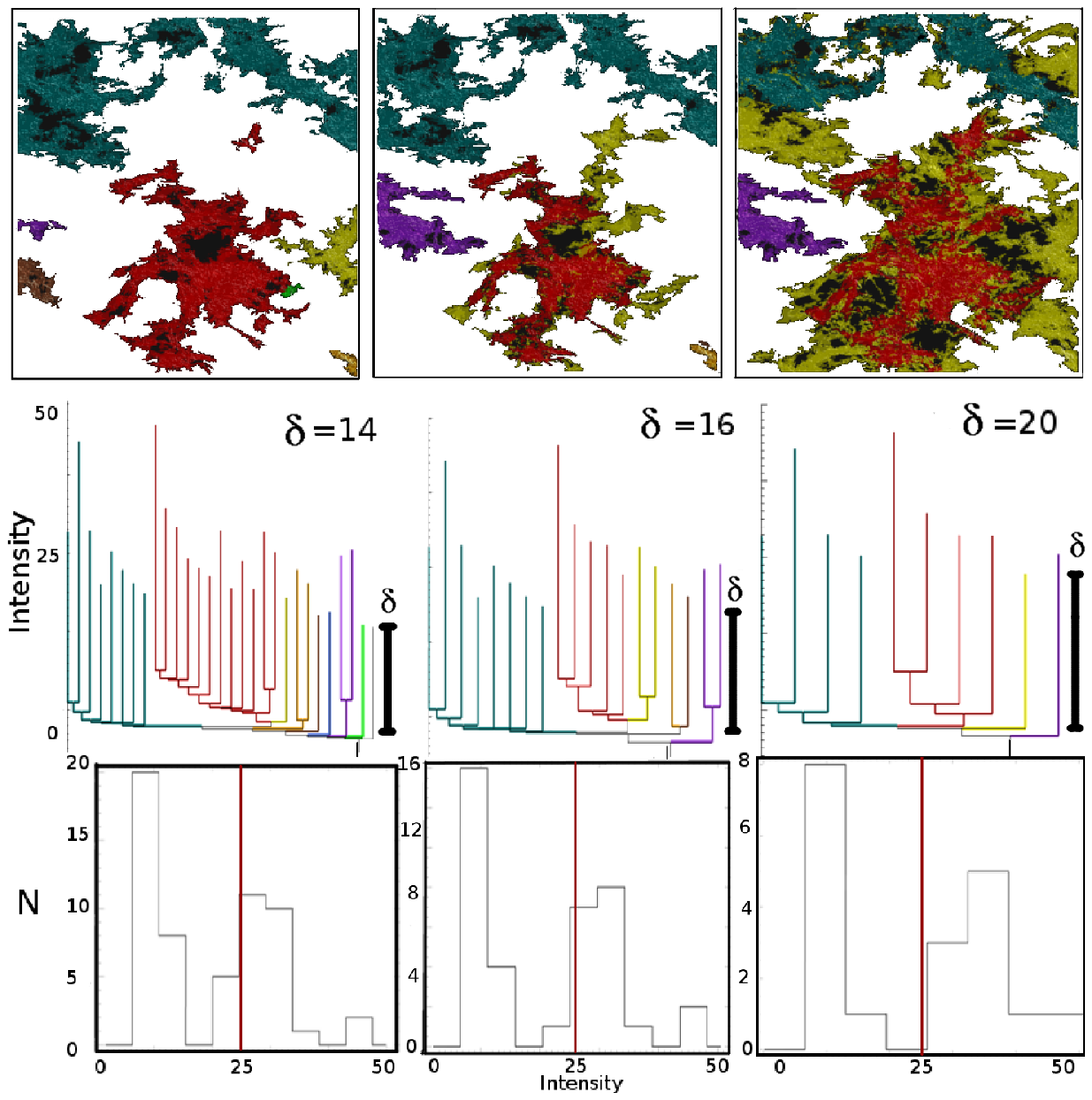}
\caption{Different ways of viewing dendrogram information used in this work.  Here we show an example for supersonic sub-Alfv\'enic turbulence
(Run 3 from Figure \ref{descrp}) for threshold values $\delta$=14, 16, 20 (left, center, right columns). 
The top row represents the isosurfaces  in the PPV data and the middle row is the corresponding dendrogram (the black line is a reference marker for $\delta$)
with colors matching to the isosurface structures. Note that there is no information on the $x$-axis of the tree diagram as the branches are sorted not to cross.  However, this still preserves
all information about connectivity and hierarchy at the expense of positional information.
The bottom row is the histogram of the resulting tree diagram, including the leaves, branches and nodes. The red
line is a reference marker at intensity level 25.
The units of intensity on the $y$-axis of the tree diagrams in the middle
row could be in brightness temperature ($T_b$) for scaled simulations or observations.}
\label{fig:hists}
\end{figure*}

As $\delta$ sets the definition for ``local maximum,'' setting it too high will produce a dendrogram that may miss important
substructures, while setting it very low may produce a dendrogram that is dominated by noise.  
The issues of noise and the dendrogram
were discussed extensively in Rosolowsky et al. 2008.  While the dendrogram is designed to 
present only the essential features of the data, noise will mask the low-amplitude or high spatial frequency variation in the emission 
structures.  In extreme cases, where the threshold value is not set high enough or the signal-to-noise is very low, noise
can result in local maxima that do not correspond to real structure.  As a result, the algorithm
has a built in noise suppression criteria which  only recognizes structures that have 4 $\sigma_{rms}$
significance above $\delta$.   Such a criterion has been previously used in data cube analysis, as noise fluctuations will 
typically produce 1 $\sigma_{rms}$  variations (Brunt et al. 2003; Rosolowsky \& Blitz 2005; Rosolowsky et al. 2008).  

The algorithm we use is extensively described in  Goodman et al. (2009) in the Supplementary Methods section and in Rosolowsky et al. (2008), however we describe its main points here.
To produce the dendrogram, we first identify a population of local maxima as the points which are larger than all
surrounding voxels touching along the face (not along edges or corners). 
This large set of local maxima is then reduced by examining each maximum and searching
for the lowest contour level that contains only that maximum.  If
this contour level is less than $\delta$ below the local maximum, that
local maximum is removed from consideration in the leaf population. This difference in data
values is the vertical length of the ``leaves'' of the dendrogram.
Once the leaves (local maxima) of the dendrogram are established, we contour the data with a large
number of levels (500 specifically, see Rosolowsky et al. 2008; Goodman et al. 2009).
The dendrogram ``branches'' are graphically constructed by connecting the various sets of maxima at
the contour levels where they are joined (see Figure \ref{fig:alyssa} for a 1D example).  For graphical presentation, the
leaves of the structure tree are shuffled until the branches do not cross when plotting. As a result, the
$x$-axis of the dendrogram contains no information.

\begin{figure}[th]
\centering
\includegraphics[scale=.74]{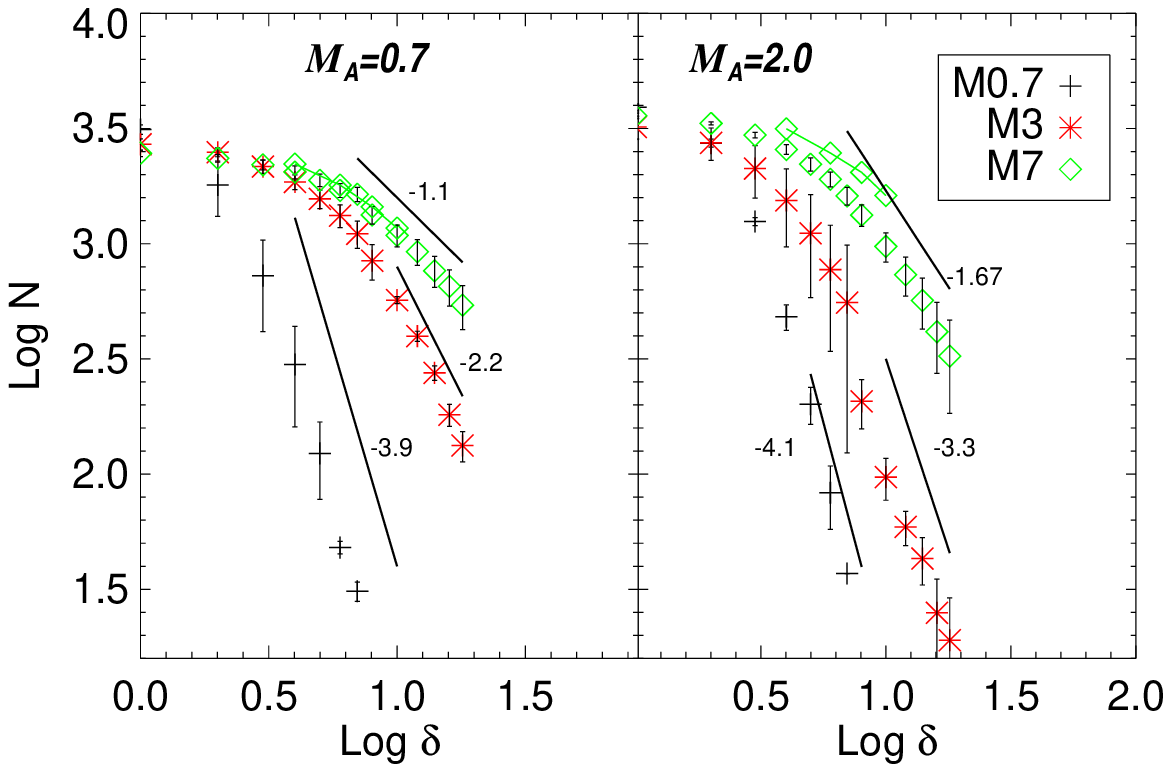}
\includegraphics[scale=.478]{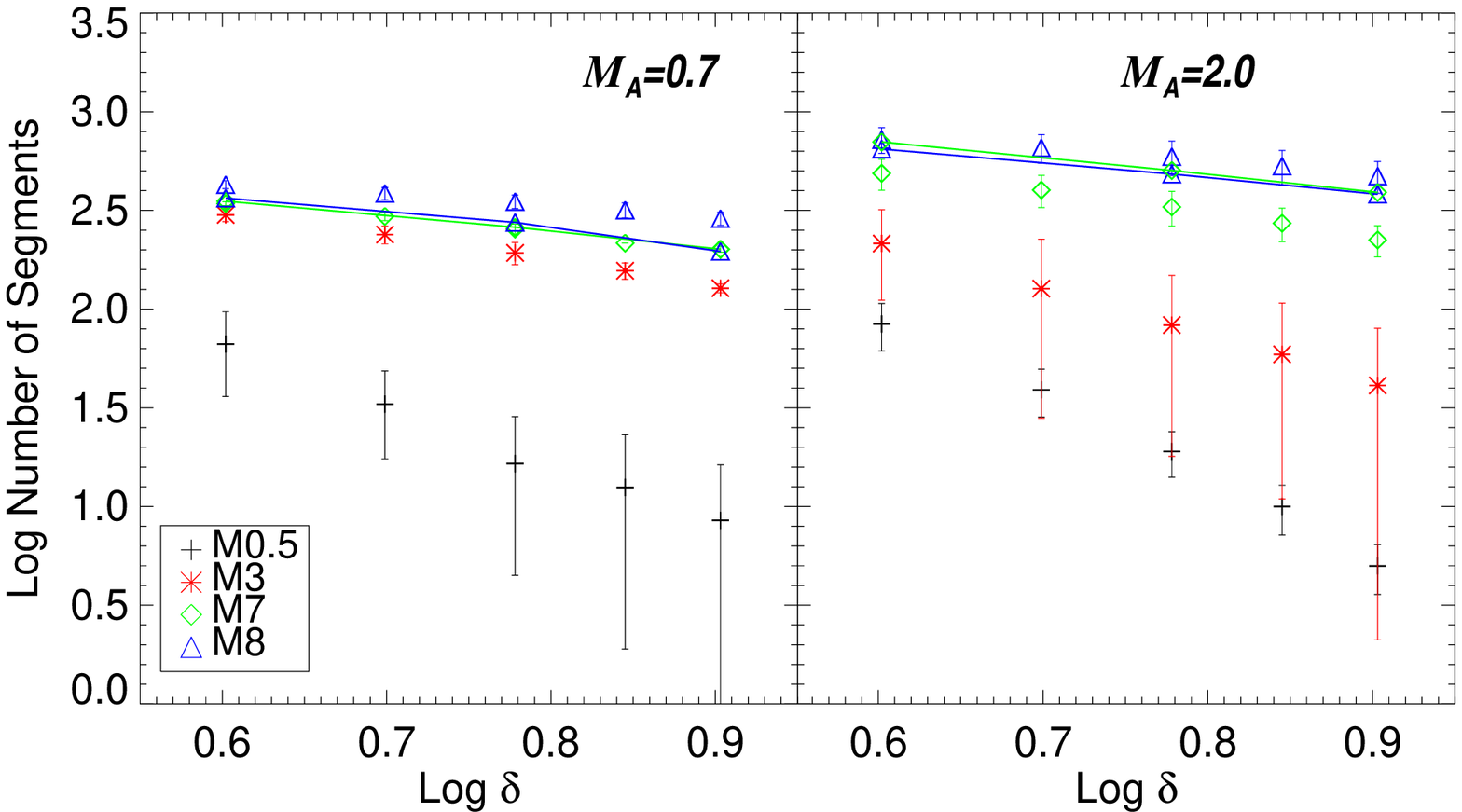}
\caption{Top: total number of structures (leaves and nodes) vs.~$\delta$  for six different simulations. 
Error bars are created by running the analysis for multiple time snapshots of well-developed turbulence. 
Symbols with no connecting line have LOS taken perpendicular to the mean magnetic field.
We show an example of a case with LOS parallel to the mean field for the M7 runs denoted with points connected with a straight line running.
Bottom: number of segments from root to leaf on the largest branch of the tree vs.  $\delta$.  
We show an example of a case with LOS parallel to the mean field for the M7/M8 runs denoted with points connected with a straight line running.
Hierarchical structure is created both by shocks (high sonic Mach number cases)  and a high  Alfv\'enic Mach number.
For both the top and bottom plots, the left panel shows higher magnetization (sub-Alfv\'enic)  while the right shows lower magnetization (super-Alfv\'enic).  
Both panels have the $y$-axis set to the same range for ease of comparison and use a log-log scale. }
\label{fig:maxvsdelta}
\end{figure}

Once the dendrogram is created, there are multiple ways of viewing the information it provides such as:
\begin{itemize}
\item A tree diagram (the dendrogram itself).
\item 3D viewing of the isocontours and their connectivity in PPV space.
\item A histogram of the dendrogram leaf and node values (i.e. intensities), which can then be further statistically analyzed.
\end{itemize}

We note that this third point is a novel interpretation of the dendrogram that we develop in this work. Here, the histogram will be composed of
intensity values important to the hierarchical structure of the image. We define a distribution $\xi$ which includes the intensity values of the leaves (i.e. local maximum), denoted by $m$, and intensity values of the nodes, denoted with $n$.  This interpretation is visualized in Figure \ref{fig:hists} and is further described in Section \ref{results}.

The purpose of this paper is to use dendrograms to characterize the observed hierarchy seen in the data.
While turbulence has often been cited as the cause of the observed hierarchical structure in the ISM (Stutzki 1998), it is unclear to what extent
magnetic fields, gas pressure, and gravity play roles in the creation of ISM hierarchy even though these parameters are known to drastically change
the PDF and spectrum of both column density and PPV data (see Falgarone 1994; Kowal, Lazarian \& Beresnyak 2007; Tofflemire et al. 2011).

\section{Data}
\label{data}
We generate a database of sixteen 3D numerical simulations of isothermal compressible (MHD)
turbulence by using the MHD code of Cho \& Lazarian 2003 and vary the input 
values for the sonic and Alfv\'enic Mach number.
 We briefly outline the major points of the numerical setup.

The code is a third-order-accurate hybrid essentially 
non-oscillatory (ENO) scheme  which solves
the ideal MHD equations in a periodic box:
\begin{eqnarray}
 \frac{\partial \rho}{\partial t} + \nabla \cdot (\rho {\bf v}) = 0, \\
 \frac{\partial \rho {\bf v}}{\partial t} + \nabla \cdot \left[ \rho {\bf v} {\bf v} + \left( p + \frac{B^2}{8 \pi} \right) {\bf I} - \frac{1}{4 \pi}{\bf B}{\bf B} \right] = {\bf f},  \\
 \frac{\partial {\bf B}}{\partial t} - \nabla \times ({\bf v} \times{\bf B}) = 0,
\end{eqnarray}
with zero-divergence condition $\nabla \cdot {\bf B} = 0$, 
and an isothermal equation of state $p = C_s^2 \rho$, where 
$p$ is the gas pressure. 
On the right-hand side, the source term $\bf{f}$ is a random 
large-scale driving force.   Although our simulations are ideal MHD, diffusion is still present in the form of numerical resistivity 
acting on small scales.
The scale at which the dissipation starts to act is defined by the numerical diffusivity
of the scheme\footnote{ ENO schemes, such as the one employed here, are considered to be generally
low diffusion ones (see Liu \& Osher 1998; Levy, Puppo \& Russo 1999, e.g )} . 
However, the dissipation scales can be estimated approximately from the velocity spectra and, in this case, 
 we estimated the dissipation scale to be $k_v \approx$30 (see Kowal, Lazarian \& Beresnyak. 2007).

We drive turbulence solenoidally\footnote{The differences between solenoidal and compressive driving is discussed more in Federrath et al. (2008).  One can 
expect driving in the ISM to be a combination of solenoidal and compressive, however both types of driving will produce shocks on a range of scales, which
is what we study here.} with energy injected on the large scales.
The large eddy turnover time is given by $\approx L/\delta V$ where $\delta V$ is the RMS velocity (with fluctuations of around unity) and $L$ is the box size.  
The magnetic field consists of the uniform background field and a 
fluctuating field: ${\bf B}= {\bf B}_\mathrm{ext} + {\bf b}$. Initially ${\bf b}=0$.  The average density is unity for all simulations.
We stress that simulations (without self-gravity) are scale-free
and all units are related to the turnover time, density, and energy 
injection scale.

We divide our models into two groups corresponding to 
sub-Alfv\'enic and
super-Alfv\'enic  turbulence. 
For each group we computed several models with different values of 
gas pressure (see Figure \ref{descrp}) falling into regimes of subsonic and supersonic.  
We ran one compressible MHD turbulent model with self-gravity at 512$^3$ resolution and a corresponding case for the same initial values of pressure 
and mean magnetic field without self-gravity (models 15 and 16).

We solve for the gravitation potential ($\Phi$) using a Fourier method similar to that described in Ostriker et al. 1999.  In this case, Equation 2 now has a
$-\rho \nabla \Phi$ term on the right hand side.  The gravitational kernal used to provide a discrete representation of the Poisson equation is:
\begin{equation}
\phi_k=2\pi G \rho_k\{[1-cos(k_x\Delta x)]/\Delta x^2+[1-cos(k_y\Delta y)]/\Delta y^2+[1-cos(k_z\Delta z)]/\Delta z^2\}^{-1}
\end{equation}

We can set the strength of self-gravity by changing the physical scaling of the simulations, i.e. by changing the size scaling, cloud mass, and crossing time, which effectively changes the relation between
$2\pi G$ in physical units and code units, which we term $g$ in Figure \ref{descrp}.
We set these values to give a global virial number of $\alpha \approx 90$, which is in the upper limit of the observed values found in GMCs (see Kainulainen et al. 2011).   We choose a high virial value to investigate the minimum effect gravity might have on the hierarchical structure of clouds.
More information about the scalings and their relation to the virial parameter can be found 
in Section \ref{sec:app}.

The models are listed and described in Figure \ref{descrp}. Here we list the root-mean-squared values of the sonic and Alfv\'enic Mach numbers
calculated in ever cell and then averaged over the box.

We use density and velocity perpendicular to the mean magnetic field in order to create fully optically thin
synthetic PPV data cubes, although we also investigate dendrogram for 
other LOS orientations.  The PPP and  synthetic PPV cubes are all normalized by the mean value, i.e.
$ PPV_{final}=PPV_{orignial}/<PPV_{orignial}>$, unless otherwise stated.
 Varying the optical depth will be done in a later work.  
 We create cubes with a given velocity resolution of 0.07, which is approximately ten times smaller than the rms velocity of the simulation ($v_{rms}\approx$ unity).
For reference, the sound speed of the simulations varies from $c_s=1.4-0.07$ for our most subsonic to most supersonic simulations. 
PPV cubes are created by reorganizing
density cubes into channel bins based on given velocity intervals.  Additional discussion
on comparing the simulations to observations is found in Section \ref{sec:app}.

\section{Characterizing Hierarchy and Structures Created by Turbulence}
\label{results}

We applied the dendrogram algorithm on synthetic PPV cubes with various sonic and Alfv\'enic Mach numbers. An example
of how the the tree diagram output changes with threshold value $\delta$ is shown in Figure \ref{fig:hists}.
The top row of Figure \ref{fig:hists} shows the isosurfaces with the colors relating back to the colors in the corresponding dendrogram shown in the middle row.
As the threshold intensity value $\delta$ (which, shown here with a black line, sets the definition of the local maximum or ``leaves of the tree'')
increases, structures in the dendrogram begin to merge with each other.  
The leaf and branch length and number of structures provide information on the hierarchical nature of the PPV cube. 
The bottom row of  Figure \ref{fig:hists} shows the histograms of the dendrogram distribution of intensities (leaves and nodes). The red line is a reference line at 
intensity level 25.  This distribution also changes with changing threshold value, as 
leaves merge with one another and the hierarchy changes.

\begin{figure*}[th]
\centering
\includegraphics[scale=.13]{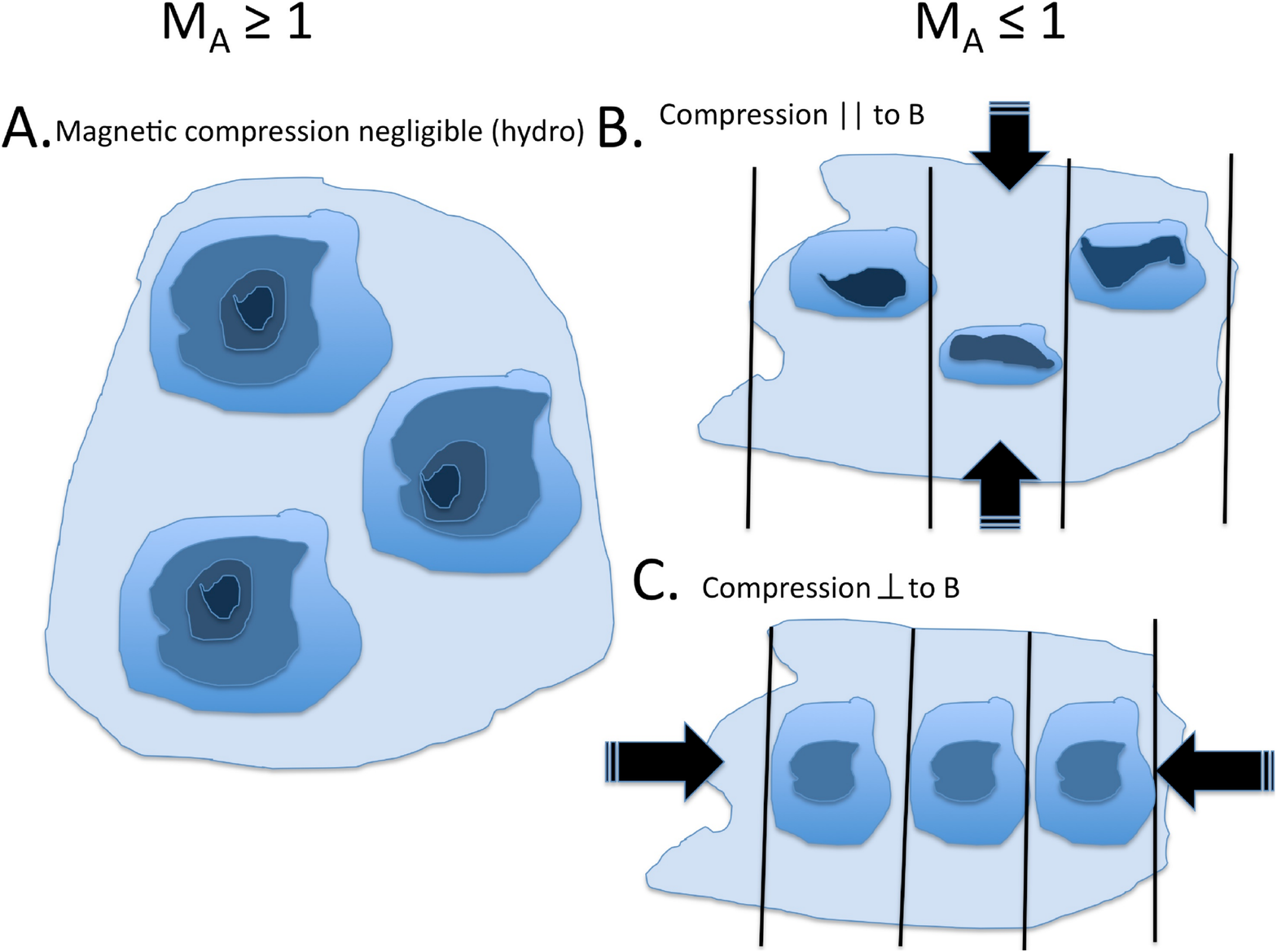}
\caption{Illustration of supersonic clouds with different magnetic regimes and how this affects the observed clumps. Panel A shows a cloud with very low Alfv\'enic Mach number  (similar to hydrodynamic turbulence). In this case, turbulence allows the creation of hierarchical structure with no limitation on the gas motion.
 Panels B and C show a different cloud, with higher magnetization (sub-Alfv\'enic), with views of the compression parallel (panel B) and perpendicular (panel C) to the field lines. In the sub-Alfv\'enic cloud, motions will
be correlated due to the strong field. The magnetic field will restrict shock compression perpendicular to the field lines (panel C).  For shocks parallel to the field (panel B), 
increased compression will occur which will enhance density clumps.}
\label{fig:clump}
\end{figure*}

In the next subsections, we investigate the effects of the compressibility, magnetization, and self-gravity on the number of structures, amount of hierarchical structure, and moments of the dendrogram 
distribution. We define a hierarchical dendrogram as one which has many segments on its paths and hence many levels above the root.

\subsection{Sonic and Alfv\'enic Mach Numbers}

\subsubsection{Leaf and Branch Counting}

We computed the dendrogram for all synthetic non-self gravitating PPV cubes with varying threshold values.
Figure \ref{fig:maxvsdelta} top shows how the 
total number of structures (i.e., dominant emission contours including dendrogram ``leaves and branches'')
 changes as we change $\delta$. 
We plot the total number of structures vs.~$\delta$ on a logarithmic scale (i.e. Log N vs Log $\delta$)
for simulations with three differing values of sonic Mach number (the M7, M3, M0.7 models)
and two values of Alfv\'enic Mach number.  The left panel shows  sub-Alfv\'enic models and the right shows super-Alfv\'enic models. 
Error bars are created by taking the standard deviation between different time snapshots. 
We note that power law tails can be seen at values of $\delta$ past the mean value (i.e. past log $\delta=0$).  We 
over plot the values of the slopes with solid black lines for reference.  
The symbols with no lines through them in Figure \ref{fig:maxvsdelta} are for PPV cubes with LOS taken perpendicular to the mean magnetic field.  We tested our results for LOS taken parallel to the 
mean magnetic field and found similar results.  We show examples 
of the results for cubes with LOS taken parallel to the 
mean magnetic field in Figure \ref{fig:maxvsdelta} for the M7 models using symbols connected with solid lines. The M7 models are highly supersonic and therefore will show the most deviation along different sight lines.

When $\delta$ is at or slightly above the mean value of the data cubes, there is little difference
in the number of structures between simulations of different sonic Mach number.
This is surprising, since the structures seen in subsonic turbulence are very different from the supersonic case.  
In the regime where $\delta$ is at the mean value,  we are sampling most of the PPV cube emission and therefore are not sensitive to the differences seen at larger threshold values, which
will merge low intensity structures.
Once we increase
$\delta$ beyond the mean however, the number of structures between the subsonic (black plus signs) and supersonic simulations (red stars and green diamonds) rapidly diverges. The larger number of structures in the supersonic case is a result of  shocks creating higher intensity values in the PPV cubes. 
Additionally, the slopes in the subsonic simulations are much steeper as compared with the  supersonic simulations since the
number of structures the dendrogram considers significant at a given threshold value rapidly falls off to zero.
Subsonic models have fewer significant emission contours
since they do not have density enhancements created by shocks and therefore
the density/velocity contrast between subsonic and supersonic turbulence becomes clear at higher threshold values.  
The higher the Mach number, the more small scale intensity enhancements we expect to see.

As $\delta$ increases,  differences between supersonic (M3) and very supersonic (M7) cases become more apparent, as the slopes
for the M3 case are steeper.  This is because interacting shocks in the M7 case are much stronger, and hence there is more contrast in the
emission contours.  Thus, as we increase $\delta$, the structures merge more rapidly for lower values of the sonic Mach number.

Comparison between the left and right top panels of Figure \ref{fig:maxvsdelta} shows that the magnetic field also affects the number of structures and the trend with the threshold value.
When $\delta$ is low, the super-Alfv\'enic cases (right panel) show slightly more structures than the sub-Alfv\'enic ones (left panel).  However, as $\delta$ increases, the number of structures decreases more rapidly in the case of super-Alfv\'enic turbulence, i.e. the slopes are steeper.  
Interestingly, when we take a sight-line parallel to the mean magnetic field (green symbols with the solid
line drawn through) we still see the same trend with Alfv\'enic Mach number.
We suspect that these results are generally independent of time evolution in driven turbulence, as clump lifetimes have been shown to be determined by turbulence motions on large scales, rather then limited by their crossing times (see Falceta-Gon\c{c}alves \& Lazarian 2011)

\begin{figure}[th]
\centering
\includegraphics[scale=.7]{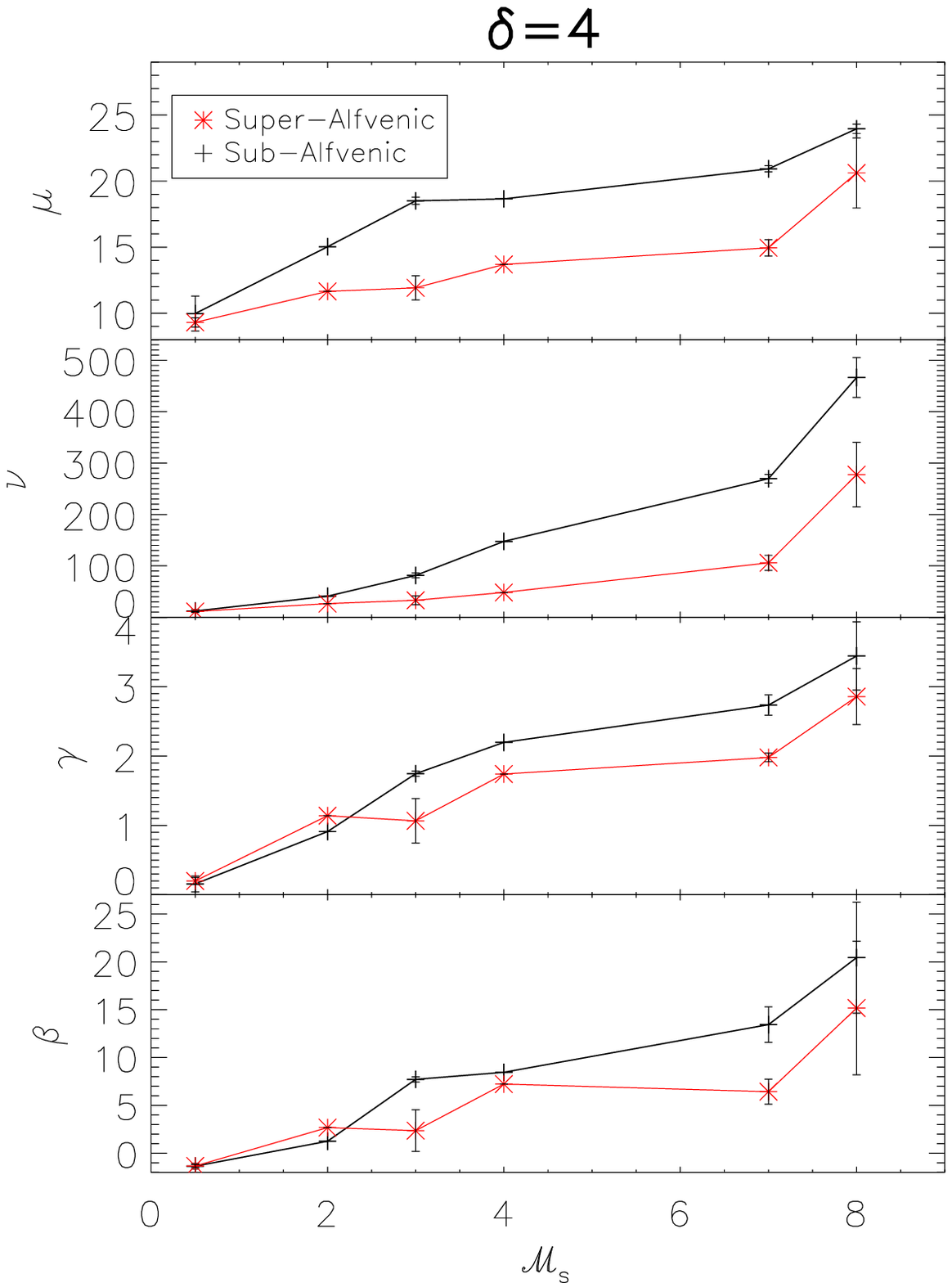}
\caption{Moments of the dendrogram tree (leaves + branches) vs. average ${\cal M}_s$ for twelve different simulations spanning a range of
sonic numbers from 0.5 to 8.  Here, we have chosen $\delta$=4.
 Panels, from top to bottom, show  mean ($\mu$), variance ($\nu$), skewness ($\gamma$) and kurtosis ($\beta$) of the distribution. 
 Sub-Alfv\'enic is shown with black plus signs and super-Alfv\'enic with red asterisks.}
\label{fig:norm_ms}
\end{figure}

The bottom plot of Figure \ref{fig:maxvsdelta}
shows the  number of segments from root to leaf on the largest branch vs. the threshold parameter $\delta$.  
A test of hierarchy is to count the number of segments along the largest branch, from leaf to root. 
Similar to what was shown  in the top figure, the sonic Mach number has a strong relation to the amount of hierarchical structure created in the gas.
Higher sonic Mach number yields more shocks, which in turn produce more high density clumps and more hierarchical structures in PPV space. 
Interestingly, the magnetic field seems to play an even stronger role in the hierarchical branching than shocks.  
Comparison between the $y$-axis values of the left and right plots reveals that 
a larger Alfv\'enic Mach number allows for more hierarchical structure in the PPV dendrogram.   
In the case of super-Alfv\'enic turbulence, magnetization is low and hence the structures created are closer to that of hydrodynamic turbulence, which is well known to show
fractal behavior and hierarchical eddies.  As turbulence transitions to sub-Alfv\'enic, it becomes magnetically dominated with fewer degrees of freedom.

We illustrate the results of  Figure \ref{fig:maxvsdelta} as a cartoon shown in Figure \ref{fig:clump}.  The cartoon shows various configurations of two ISM clouds.  We label the first cloud as case A, which is a cloud with a global Alfv\'enic Mach number $\geq 1$.  The second cloud, which is a cloud with a global Alfv\'enic Mach number $\leq 1$,  is labeled as cases B and C, indicating shock compression parallel and perpendicular to the mean magnetic field, respectively.
Both clouds are assumed to have the same supersonic value of the sonic Mach number.  Case A shows hierarchical structure
forming in clumps that are not affected strongly by the magnetic field. The clumping and 
hierarchy is due to compression via shocks and the shredding effect of hydrodynamic turbulence.  
The turbulent eddies for cloud A can evolve with a full 3D range of motion and have more degrees of freedom as compared with turbulence in the presence of a strong magnetic field. 
In light of this, when we consider strong magnetization, we must now investigate the effects of shock compressions oriented parallel and perpendicular to the mean magnetic field (case B and C).  For shock compression parallel to the field lines (case B), the clumps will be confined in the direction
perpendicular to the field, and thus the compression will be able to squeeze the clumps, decrease the hierarchy in the gas, and create additional large density contrast.
For shock compression perpendicular to the field lines, the magnetic pressure relative to the shock compression is  higher, and the 
clumps will not feel as much of the compression.  
Furthermore, the strong field creates anisotropy  in the eddies, which are stretched along the direction of the mean field line. This limits the range of motion of the eddies, which in turn limits their ability to interact.
Thus, in comparison of  cloud B/C with cloud A  the contrast is higher while hierarchical structure is less.

\subsubsection{Statistics of the Dendrogram Distribution}
\label{moments}

A dendrogram is a useful representation of PPV data in part because there are multiple ways of exploring the information on the data hierarchy.  
In this section we  investigate how the statistical moments of the 
distribution of the dendrogram tree (see bottom panels of Figure \ref{fig:hists} for example) changes as we change the threshold parameter
$\delta$ and how these changes depend on the compressibility and magnetization of turbulence.  We consider a distribution  $\xi$ containing 
all the intensity values of the leaves and merging intensity contour values in a given dendrogram.  The question that forms the basis of our investigation in this section is:
Do the moments of the distribution $\xi$ have any dependencies on the sonic and Alfv\'enic Mach numbers?

The first and second order statistical moments (mean and variance) used here are defined as follows:
$\mu_{\xi}=\frac{1}{N}\sum_{i=1}^N {\left( \xi_{i}\right)} $
and $ \nu_{\xi}= \frac{1}{N-1} \sum_{i=1}^N {\left( \xi_{i} - \overline{\xi}\right)}^2$, respectively. 
The standard deviation is related to the variance as: $\sigma_{\xi}^2=\nu_{\xi}$.
The third and fourth order moments (skewness and kurtosis) are defined as:

\begin{equation}
\gamma_{\xi} = \frac{1}{N} \sum_{i=1}^N{ \left( \frac{\xi_{i} - \mu_{\xi}}{\sigma_{\xi}} \right)}^3 
\label{eq:skew}
\end{equation}

\begin{equation}
\beta_{\xi}=\frac{1}{N}\sum_{i=1}^N \left(\frac{\xi_{i}-\mu_{\xi}}{\sigma_{\xi}}\right)^{4}-3
\label{eq:kurt}
\end{equation}

We calculate the moments of the dendrogram tree distribution while varying our simulation parameter space. In particular, we
vary the sonic Mach number, the Alv\'enic Mach number, and the threshold value.
We find the moments vs. the threshold parameter $\delta$ to show linear behavior.
As $\delta$ increases, the number of the intermediate intensity values that make up the branches and the hierarchical nesting (i.e. 
the intensity values between the high intensity local maximum and the low intensity values near the trunk)
merge with each other.  This effect can be seen visually in Figure \ref{fig:hists}.  Thus, as $\delta$ increases, the mean and variance
of the distribution $\xi$ (example show in the bottom of Figure \ref{fig:hists}) will increase.
 
We plot the moments vs.  ${\cal M}_s$ with  $\delta=4$ in Figure \ref{fig:norm_ms}.
This figure shows  the full range of our simulations with sub- and super-Alfv\'enic combinations.    
Generally, as the sonic Mach number increases so do the moments.  We found this trend to be consistent over a range of $\delta$ values, and hence only plot one case here. 
Error bars, created by taking the standard deviation of the value between different time snapshots of the simulation,
generally increase with sonic number as the 
fluctuations become increasingly stochastic and shock dominated.  
The increase of the moments of $\xi$ is related to the compressibility of the model and
more supersonic cases display more prominent clumpy features, which drive up both the average and the variation from average.  
The tails and peak of the distribution
also become increasingly skewed and kurtotic towards higher values of intensity and the distribution becomes increasingly peaked around the mean value.

It is interesting
to note that a strong dependency on the magnetization of the model exists, particularly as the sonic number goes up.  
The sub-Alfv\'enic simulations show increased moments, which implies that they exhibit more contrast (mean value is higher) and more
skewed/kurtotic distributions in their gas densities.

In the above analysis, the distribution $\xi$ included all leaves and branches of the  dendrogram tree.  
We could further cut the tree into its respective branches and leaves and analyze the distributions separately, which provides
additional constraints on the parameters.  We investigated the statistical moments on the histograms of
the branch lengths, leaf lengths, and leaf intensities and found the trends discussed above to be consistent with the results
of Figure \ref{fig:norm_ms}, and hence do not include the plots.

\subsection{Self-Gravity}
\label{sg}
\subsubsection{Leaf and Branch Counting}
\label{sglb}

The issues of the importance of self-gravity in simulations for comparisons with the molecular medium have been raised by a number of authors 
(Padoan et al. 2001; Li et al. 2004; Goodman et al. 2009; Federrath et al. 2010). 
While self-gravity is known to be of great importance to accretion disk physics and protostellar collapse, its role in the outer regions of GMCs and in 
diffuse gasses is less obvious.
The dendrogram can potentially be used to explore the relative role between turbulence and gravity regarding 
both the structure of the hierarchy and the distribution of dominant emission contours.   

\begin{figure*}[tbh]
\centering
\includegraphics[scale=.34]{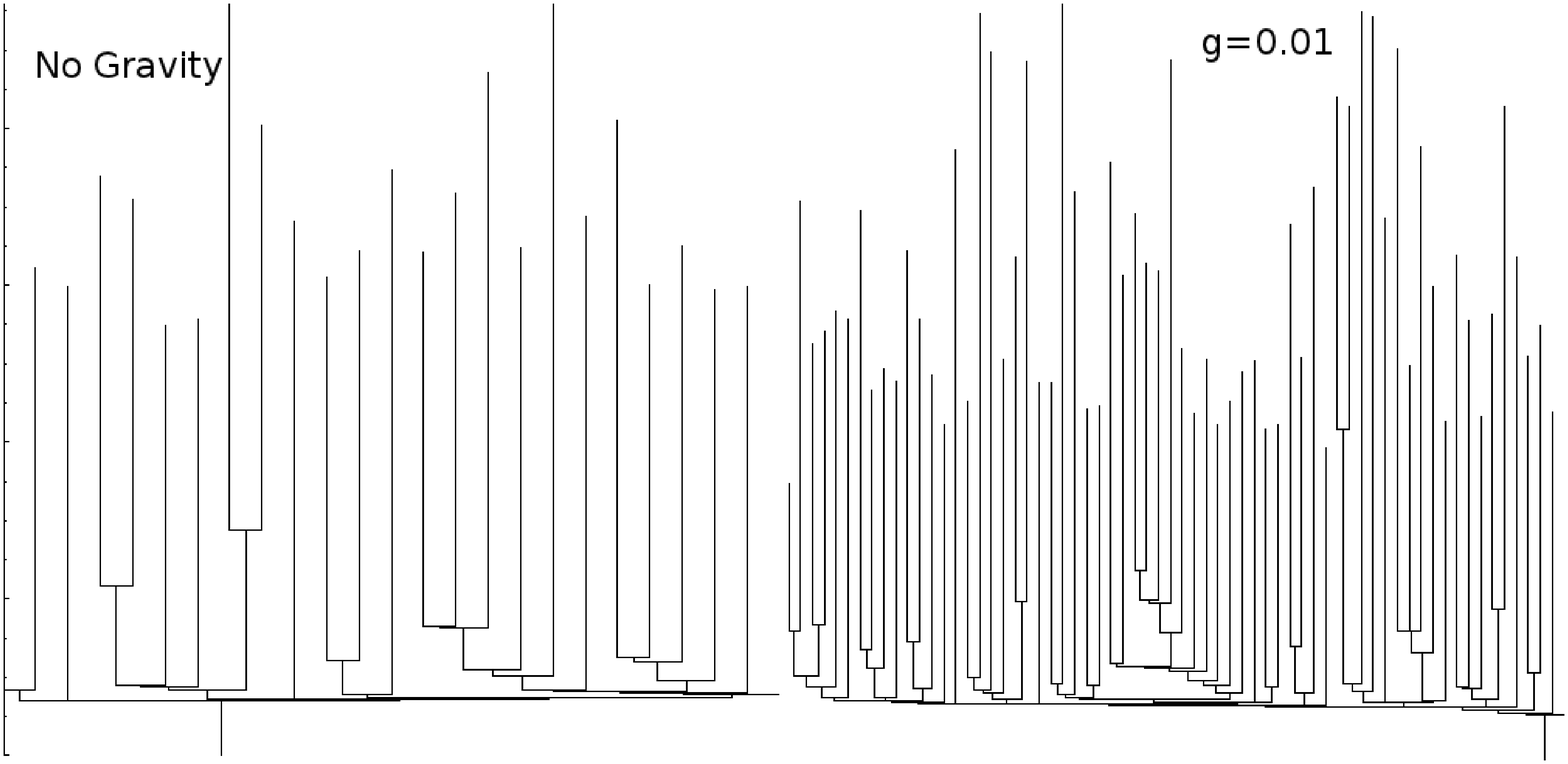}
\caption{Effects of self-gravity on the dendrograms of sub-Alfv\'enic supersonic MHD turbulence with $\delta=45$.
A high value of $\delta$ is used to keep the plots from being over crowded with branches.
The dendrogram of the self-gravitating simulation is on the right while the dendrogram of the simulation with no gravity on the left.}
\label{fig:sg1}
\end{figure*}

Figure \ref{fig:sg1} shows tree diagrams at constant threshold value $\delta$=45 for sub-Alfv\'enic supersonic simulations with and without self-gravity (models 15 and 16 in Figure \ref{descrp}). 
A large value of $\delta$ is used in order to not over crowd the dendrogram with branches.
It is clear that the case with self-gravity (the tree diagram in the right panel of Figure \ref{fig:sg1}) has a dendrogram with more significant structure and hierarchical nesting. 
Both models are supersonic with approximately the same sonic Mach number. Visually, Figure \ref{fig:sg1} makes it clear that self-gravity, even with a virial number of 90, is an important factor
for the creation of hierarchical structure.  

\begin{figure}[tbh]
\centering
\includegraphics[scale=.5]{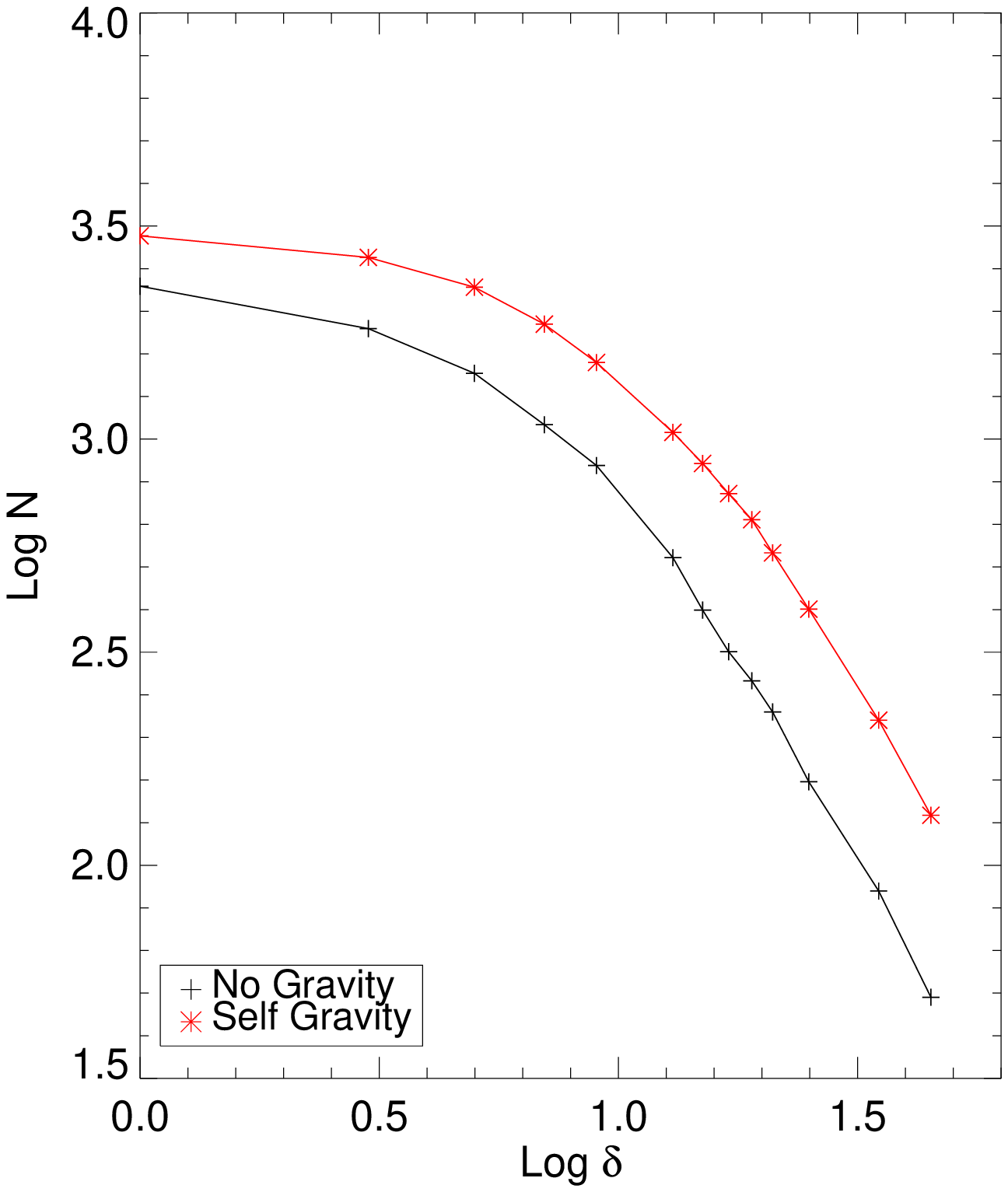}
\includegraphics[scale=.5]{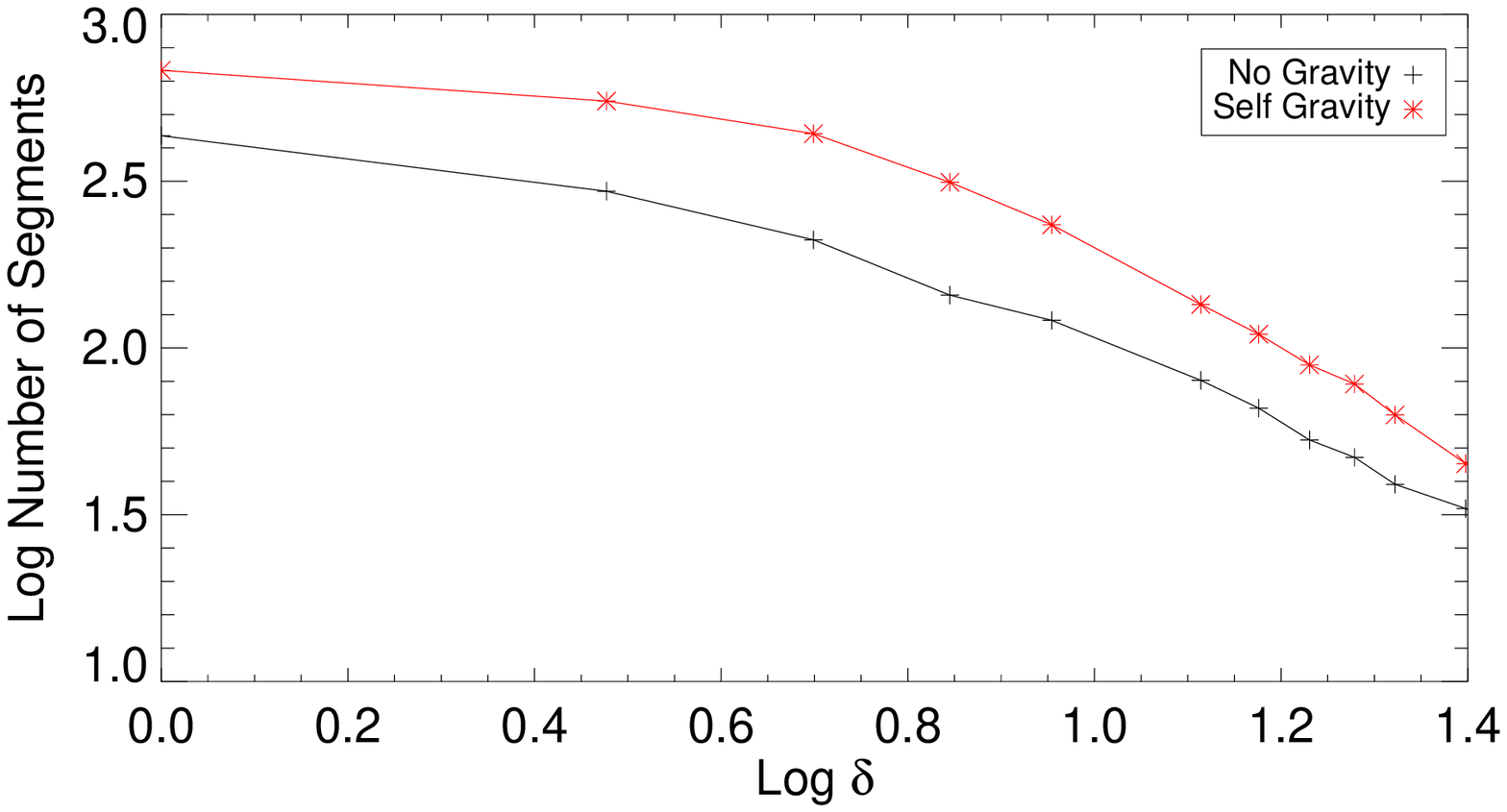}
\caption{The effects of self-gravity on the number of structures and hierarchical nesting seen in the dendrogram.
Top plot: total number of structures
(branches and leaves) vs. $\delta$.
Bottom plot: number of segments from root to leaf on the largest branch of the tree vs.  $\delta$.    All plots are shown with a log-log scale.
Black plus signs indicated the simulation with no gravity while red asterisk indicate the simulation with self-gravity included.  Both 
simulations are sub-Alfv\'enic and supersonic.}
\label{fig:sg5}
\end{figure}

In order to quantify the number of structures and hierarchical nesting seen in Figure \ref{fig:sg1} we repeat the analysis performed in Figure \ref{fig:maxvsdelta}.
We show the number of structures vs. $\delta$ (top panel) and the number of segments on the largest branch of the tree vs.  $\delta$ (bottom panel) in    Figure \ref{fig:sg5}. 
It is clear that the case with no self-gravity (black crosses) shows less overall structure (leaves and nodes, top plot) and less hierarchical structure (bottom plot)
compared with the cases with self-gravity (red asterisk).
Even with a high virial parameter, the supersonic self-gravitation simulation has significantly more nested structures and more contours considered to be 
areas of significant emission than supersonic simulation without gravity.  Interestingly, the power-law slopes seen in the top plot of Figure \ref{fig:sg5} 
are not significantly different between the self-gravitating and non self-gravitating cases.  In fact, the simulations appear to be only shifted vertically and no substantial difference is seen regarding the slopes.  Therefore it is possible that the change in the slope, which was observed in Figure \ref{fig:maxvsdelta},
can be attributed to the sonic and Alfv\'enic nature of the turbulence directly, and is not substantially affected by the inclusion of  weak self-gravity.  More work should be done in the future to determine how the dendrogram is influenced by the presence of strong self-gravitating flows.
We discuss further the use of the dendrogram for quantifying the relative importance of self-gravity and turbulence in the discussion section (section \ref{disc}).

\subsubsection{Statistics of the Dendrogram Distribution}
\label{sgstat}

\begin{figure}[tbh]
\centering
\includegraphics[scale=.7]{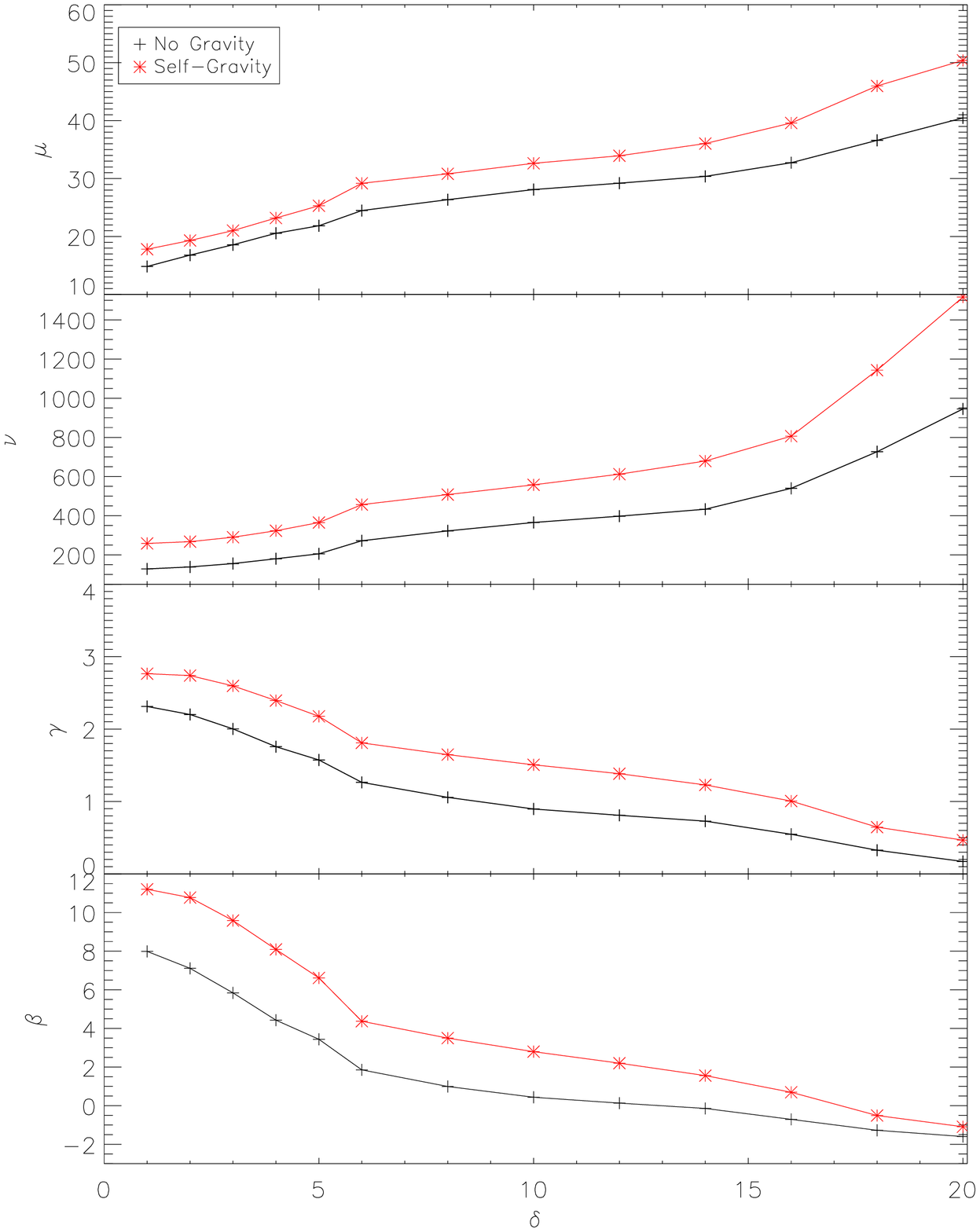}
\caption{The effects of self-gravity on the moments of the dendrogram distribution vs. $\delta$. 
Self-gravity (red asterisk) shows increased mean ($\mu$), higher variance ($\nu$), and more skewed and peaked distributions, which are
reflected in the skewness ($\gamma$) and kurtosis ($\beta$).}
\label{fig:sg6}
\end{figure}

We show how self-gravity affects the moments of the dendrogram distribution as we vary $\delta$ in Figure \ref{fig:sg6} for models 15 and 16.  
Higher levels of self-gravity show increases in all four moments  over a range $\delta$.  
The presence of gravity increases the mean value and variance of emission contours as well as skews the distribution towards
higher values due to the presence of attracting regions.  

The results of Section \ref{sg} can be applicable to molecular clouds where, given
a relatively constant value of the sonic Mach number, regions under the influence of stronger self-gravity could possibly be identified with the use of 
the analysis presented in Figures \ref{fig:sg5} and \ref{fig:sg6}.  This could be done by comparing the dendrogram moments and number of structures with different $\delta$ between various sections of a cloud.
 Areas that show increased moments and increased hierarchical clumping with no changes in the slopes of structures vs. $\delta$, may indicate changes in self-gravity.  Knowing the Mach number of these regions will greatly increase the reliability of such analysis and fortunately, other techniques exist
to find these (see for example Burkhart et al. 2010; Esquivel \& Lazarian 2011; Kainulainen \& Tan 2012).    We further discuss applying the dendrogram on observations in section \ref{disc}.

\section{Dendrograms of PPP vs. PPV}
\label{ppp}
The issue of interpreting structures seen in PPV space has vexed researchers for over a decade (see Pichardo et al. 2000).  
How the structures in PPV translate to PPP depends on many factors, most importantly the nature of the turbulent environment.
The dendrogram presents a unique way of studying  how the hierarchy of structures seen in density space (PPP) relate to PPV space via  simulations.

\begin{figure*}[tbh]
\centering
\includegraphics[scale=.5]{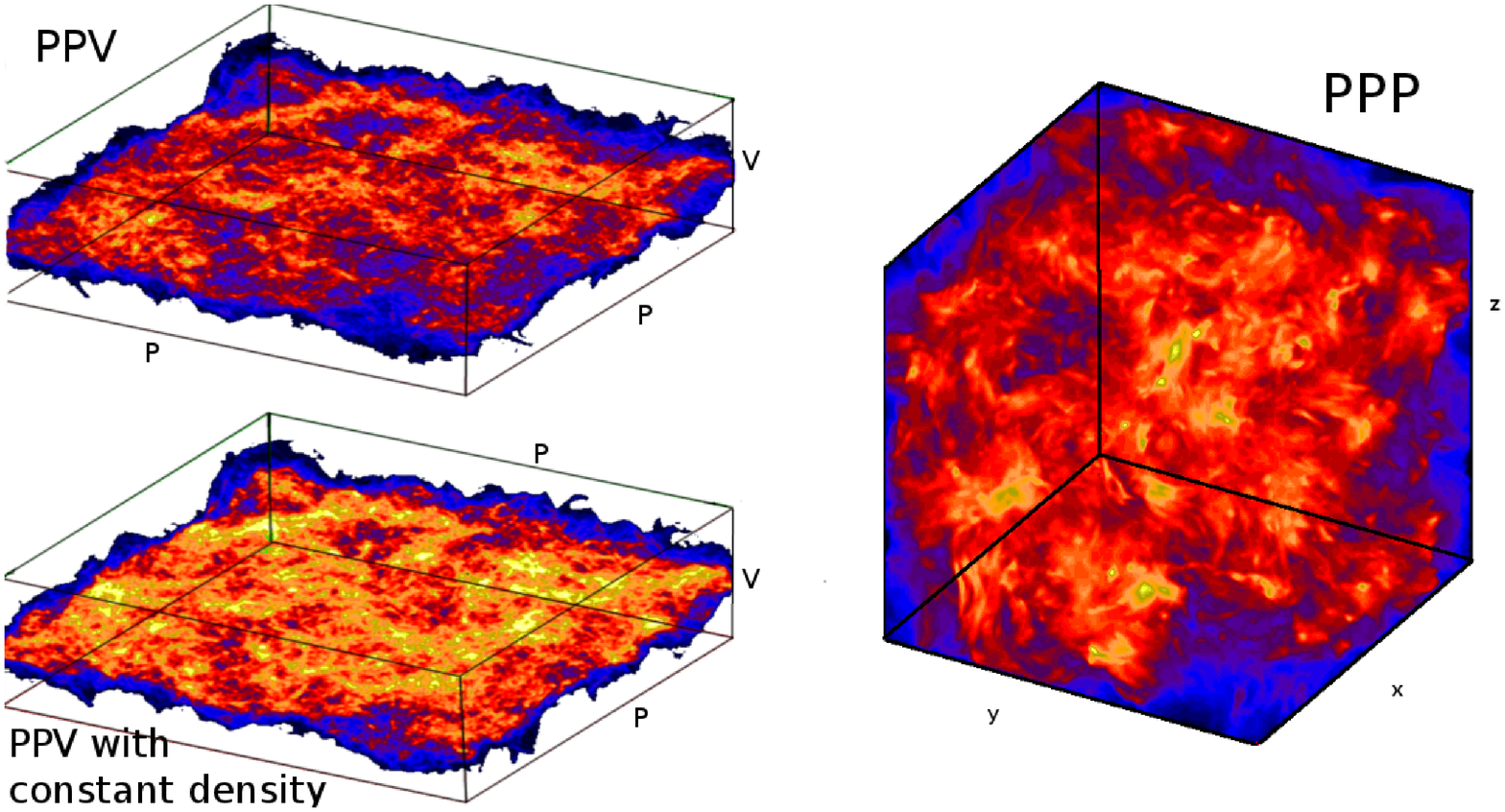}
\caption{Example of synthetic PPV data cubes with vertical axis being the velocity axis
 (left),  and PPP data cube (right)  for subsonic super-Alfv\'enic turbulence.
Integrating along the velocity axis of PPV restores the 
column density map which can also be obtained from the 3D density cube. 
The bottom left PPV has PPP density equal to unity, and hence a constant column density. Structure in this PPV cube
is due to \textit{pure velocity fluctuations}.  This figure highlights the need to be cautious when interpreting the structures seen in PPV. 
The quantitative relation between the fluctuations in PPV and underlying density and velocity fluctuations is provided in Lazarian \& Pogosyan (2000)}
\label{fig:ppv}
\end{figure*}

For turbulent clouds, it is never the case that the structures in PPV have a one-to-one correspondence with 
the density PPP, although this assumption may be  more appropriate for some environments than others.
We show a simple example illustrating this in Figure \ref{fig:ppv}, which shows two synthetic  subsonic super-Alfv\'enic
PPV data cubes (left), which share the same velocity
distribution but have different density distributions.  The bottom left PPV cube has constant density/column density, while the top left PPV cube's corresponding turbulent density cube (PPP cube) is shown on the right.

Interestingly, the bottom left PPV cube has a very similar level of structure as compared with the top PPV cube, despite the fact that the column density of the bottom
cube is constant.  This points out the well known fact that there is not a one-to-one correspondence with PPV and PPP space.
In fact, in this example (a subsonic model) most of the 
structures seen in PPV are due to the velocity rather than the density. 
Figure \ref{fig:ppv}  illustrates the dominance of velocity in the subsonic case in 
the bottom PPV cube.  Fluctuations in PPV here are \emph{entirely driven by the turbulent velocity field}.  
 
\begin{figure}[th]
\centering
\includegraphics[scale=.43]{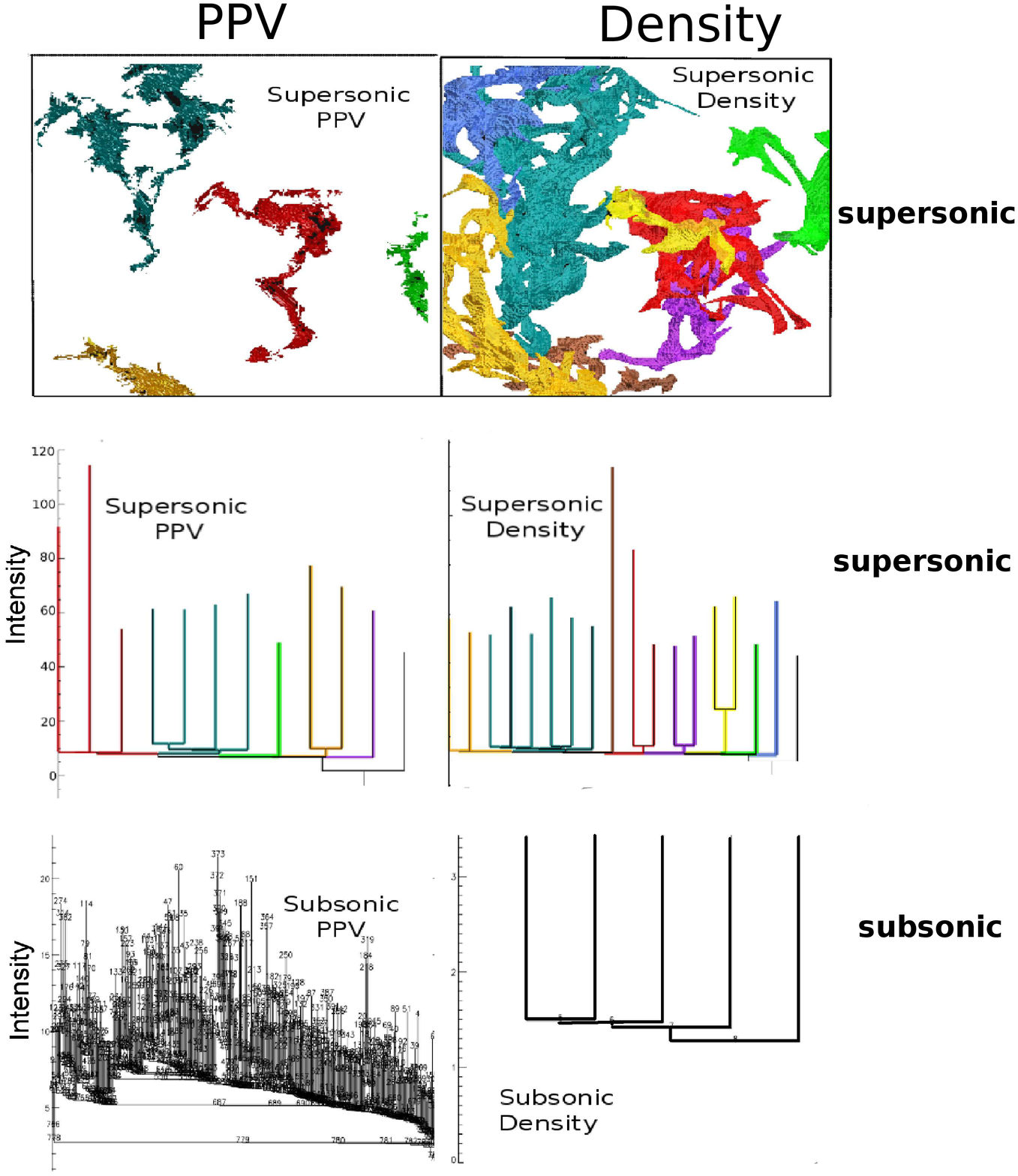}
\caption{Dendrograms of density (right column) and PPV (left column).  Supersonic isosurfaces
and their corresponding dendrograms are shown in the top and middle rows, respectively. Colors are correspondent between structures in the isosurface figures and the dendrogram.
 Subsonic dendrograms are shown in the bottom row. }
\label{fig:ppp-ppv}
\end{figure}

To illuminate this point further, Figure \ref{fig:ppp-ppv} shows PPP and PPV dendrograms for supersonic turbulence with   (model 15, middle)
and subsonic turbulence with (model 1, bottom).  We also show the corresponding isosurfaces for the supersonic case in the top row.
Comparing PPV and PPP should be done with care as they are different spaces.  We increased the value of $\delta$
until the  PPP dendrogram becomes mostly leaves, i.e. they have little hierarchy. The leaves are reached at $\approx$ $\delta =40$.   
We took the corresponding optically thin PPV cube and applied the dendrogram with the same $\delta=40$ threshold value. If the dominant
emission is due to \textit{density} than the leaves should be similar for both PPV and PPP.  All PPV and PPP
cubes are normalized to have a mean value of unity.  

 Interestingly, the supersonic turbulence dendrogram for density looks very similar to the corresponding PPV dendrogram for the same $\delta$ at the level of the leaves. 
 For the subsonic case we see that the dendrogram of density and PPV look nothing alike (same $\delta$).  In this case,
the velocity field dominates PPV space.  Hence, we do not show the isosurfaces for the subsonic case.  In supersonic turbulence, 
the highest density peaks correspond to the highest intensity fluctuation in the PPV. This implies that if one knows that the turbulence in question is supersonic, the structures in PPV space at the level of the leaves 
can generally be interpreted as 3D density structures.  However, if the turbulence is subsonic in nature this assumption is not appropriate.

\section{Application}
\label{sec:app}
The dendrogram shows dependencies on the parameters of turbulence that are important
for studies of star forming regions and the diffuse ISM. 
When analyzing a particular data set, one should keep in mind that comparisons between the observational and scaled numerical data,
or comparisons between different clouds or objects within the same data set, are the most useful means of extracting these parameters.

Our simulations can be scaled to observations by 
specifying the physical size of the simulation volume, the isothermal sound speed of the gas, and mass density.
In particular, appropriate scalings must be made in the case of the self-gravitating simulation.
We can set the strength of self-gravity by changing the physical scaling of the simulations, i.e. by changing the box size, cloud mass, and crossing time.
In this case, the relevant scaling between physical and code units are the size scaling ($x_0$), the velocity scale factor ($v_0$), 
the time scale factor ($t_0$), and the mass scale factor ($m_0$).
These are given by: 
\begin{equation}
x_0=\frac{L_{obs}}{L}
\end{equation}
where $L_{obs}$ is the physical length of the box, and $L$ is in code units and is equal to unity.
\begin{equation}
v_0=\frac{c_{s,obs}}{c_{s,code}}
\end{equation}
where $c_{s,obs}$ and $c_{s,code}$ are the observed and code sound speeds, respectively, 
\begin{equation}
t_0=\frac{x_0}{v_0}
\end{equation}
and 
\begin{equation}
M_0=\frac{M_{obs}}{\rho_{code}L^3}
\end{equation}
where $m_{obs}$ is the observed cloud mass and $\rho_{code}$ is the average density of the simulation (unity). 

Using these relations between the code units and physical units we can define a relation between $2\pi G$ 
in code units and physical units,  and then relate this to the free fall time. In this case,
\begin{equation}
2\pi G_{code}=g=2\pi G_{physical}\left({\frac{t_0}{x_0}}\right)^2\frac{M_0}{x_0}
\end{equation}

Here we use a value of  $2\pi G_{code}=g=0.01$ (see Figure \ref{descrp}).  In this case, our free fall time (in code units and therefore using $\rho$ of unity) is $t_{ff} \equiv \frac{1}{4}\sqrt{\frac{3\pi^2}{0.01\rho}}=13.6$.
The dynamical time (in code units with simulations box size of unity and rms velocity) of our self-gravitating simulation is $t_{dyn}\equiv L/v_{rms}=1.4$.
The ratio of the free fall time to the dynamical time of the simulation with self gravity  gives the global
virial parameter $\alpha\approx({\frac{t_{ff}}{t_{dyn}}})^2\approx 90$.
Additional information on scaling simulations to observations can be found in Hill et al. 2008.

We include 
the effects of changing the velocity resolution, thermal broadening, and smoothing in the next subsection.
\subsection{Smoothing}
\label{obs}

We investigate how smoothing and data resolution affect the dendrogram. When dealing with observational data, 
one must always consider the effect that the telescope beam smoothing will have on the measurement. The observations are rarely 
done with pencil beams, and the measured statistics change as the data is averaged.
We expect the effect of smoothing to depend on a dimensionless number, namely, the ratio of the size of the turbulence 
injection scale to the smoothing scale.  

We apply the same technique that was applied in the previous sections, i.e. exploring the number of structures and moments of dendrogram tree statistics.
However, we now include a boxcar smoothing kernel (truncating the edges).
We expect that smoothing will affect supersonic turbulence and cases of high self-gravity the most.  In this case, shocks and small-scale gravitational clumps become smoothed out and more difficult for the algorithm to identify.  In the subsonic or low gravity cases, smoothing makes less of a difference, since the gas is already
diffuse and less hierarchical.

\begin{figure}[tbh]
\centering
\includegraphics[scale=.7]{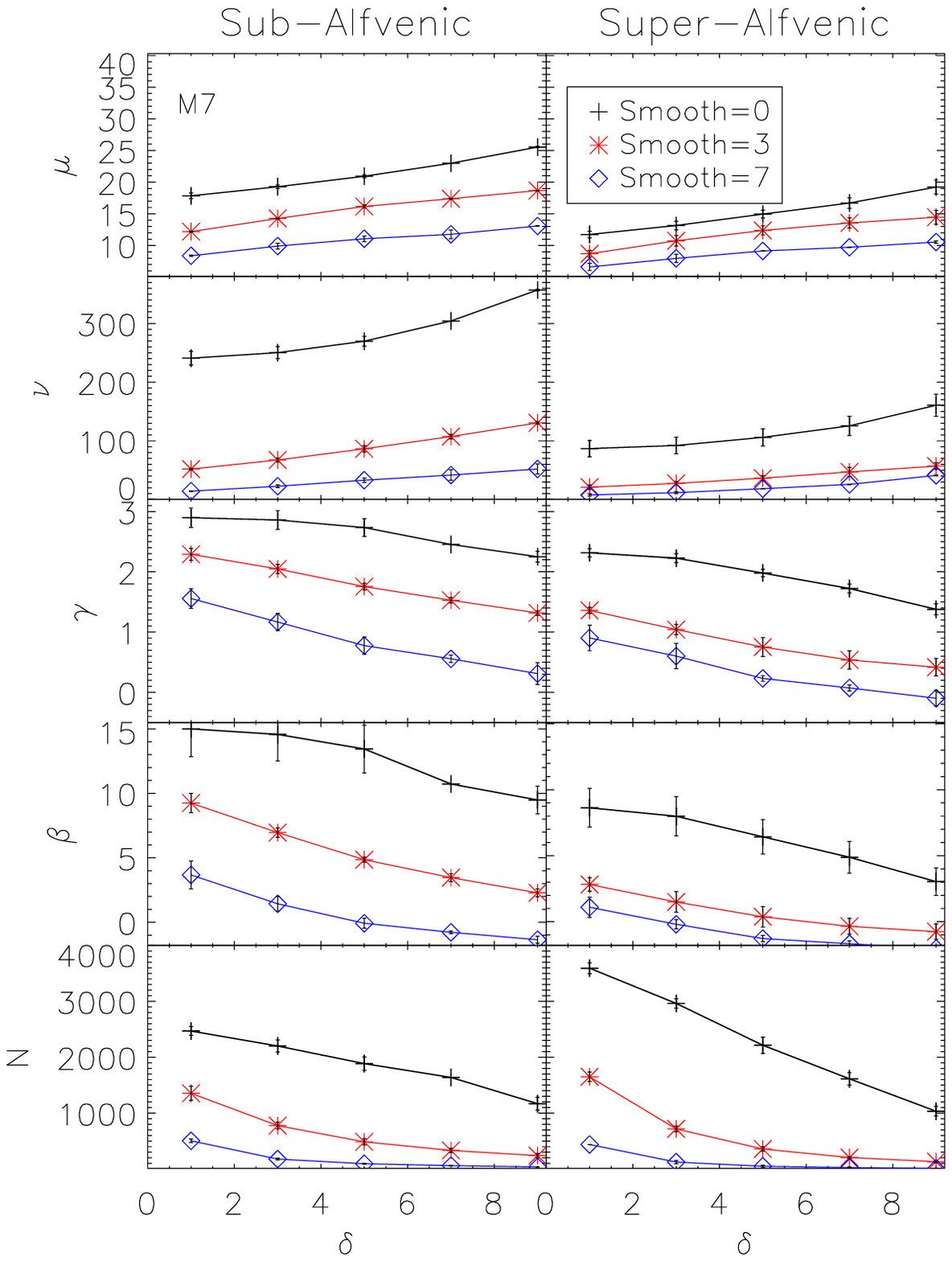}
\caption{Moments of the dendrogram distribution (top four panels) and the number of structures (bottom panel) with smoothing vs. the threshold parameter $\delta$ . 
The left panel is sub-Alfv\'enic and the right panel is super-Alfv\'enic and the $y$-axis is the same for both columns for ease of comparison between the two.
Both cases are for the M7 models.}
\label{fig:smoothmomnumb}
\end{figure}

We show how the moments and number of structures changes with  smoothing size (in pixels) in Figure~\ref{fig:smoothmomnumb}.
One could also discuss smoothing beam size in terms of the injection scale of the turbulence.  For instance, 7 pixel smoothing represents a beam scale that is
30 times smaller than our injection scale of turbulence.

We found that in general, subsonic and transonic turbulence are not as affected by smoothing compared to highly supersonic models.  In light of this, we plot
the moments and number of structures vs. $\delta$ for different smoothing degrees for a highly supersonic model with  (M7 models) in  Figure ~\ref{fig:smoothmomnumb}.
Two panels show different  Alfv\'enic regimes with the 
$y$-axis the same for both for ease of comparison.  Black lines indicate no smoothing, while red and blue indicate three and seven pixel
smoothing, respectively.  Error bars are produced by taking the standard deviation between different time snapshots of the simulations with
well developed turbulence.

As smoothing increases for this supersonic model, we see that the values of the moments as well as the total number of structures decreases.
However, even out to seven pixel smoothing the differences between the Aflv\'enic cases is evident in the mean and variance, respective of the error bars.
Furthermore, the trends with the threshold parameter do not change when we introduce smoothing, which gives us further confidence that this
technique can be applied to the observational data.  Other than the change in amplitude, the trends remain the same as what was seen in 
Section \ref{results}.

\subsection{Velocity Resolution and Thermal Broadening}

In addition to smoothing, we must also consider the effects of velocity resolution. As the velocity resolution changes
in PPV space, so do the structures observed.  
We investigated how the number of structures in the dendrogram distribution change when we vary the velocity resolution.  
We find that the number of substructures drops as the velocity resolution decreases, from several hundreds 
to several dozen when changing the velocity resolution from $v_{res}=0.07$ to $v_{res}=0.7$. 
This effect corresponds to the channel sampling dropping from $\approx$ 60 to 15 channels. This may provide
too low a number of statistics in the dendrogram distribution to look at the moments, however the general trends with 
the physical parameters stay consistent with section \ref{results}.  

An additional observational consideration that should be made regards thermal broadening.
The bulk of the this paper focuses on the effects of turbulence and magnetic fields in the creation of hierarchical structure in ISM clouds. 
However, for warm subsonic or transonic gas, thermal broadening effects should also be considered.  
Convolution with a thermal broadening profile (i.e. a Gaussian) will smooth out the velocity profiles in these cases.  However, we expect the structures seen in supersonic gas to not be affected since turbulence dominates the line broadening.  

To demonstrate this we convolve the line profiles of 
four of our simulations (models M0.7, M3, M7, M8) with Gaussian profiles
to mimic the effects of thermal broadening.  
The thermal Gaussian has FWHM given as the ratio of the turbulent line width to the sonic Mach number.

\begin{figure}[tbh]
\centering
\includegraphics[scale=.73]{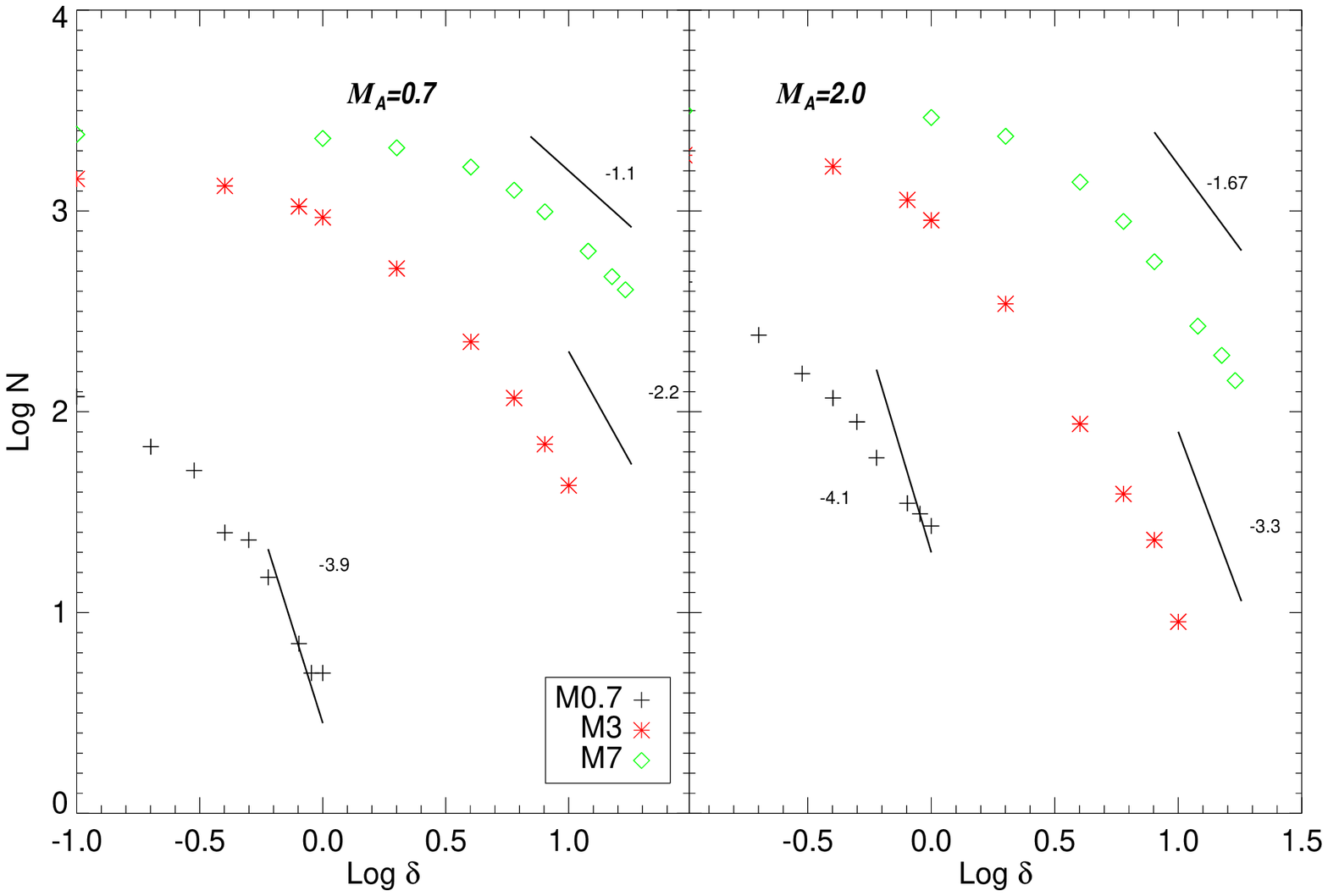}
\includegraphics[scale=.73]{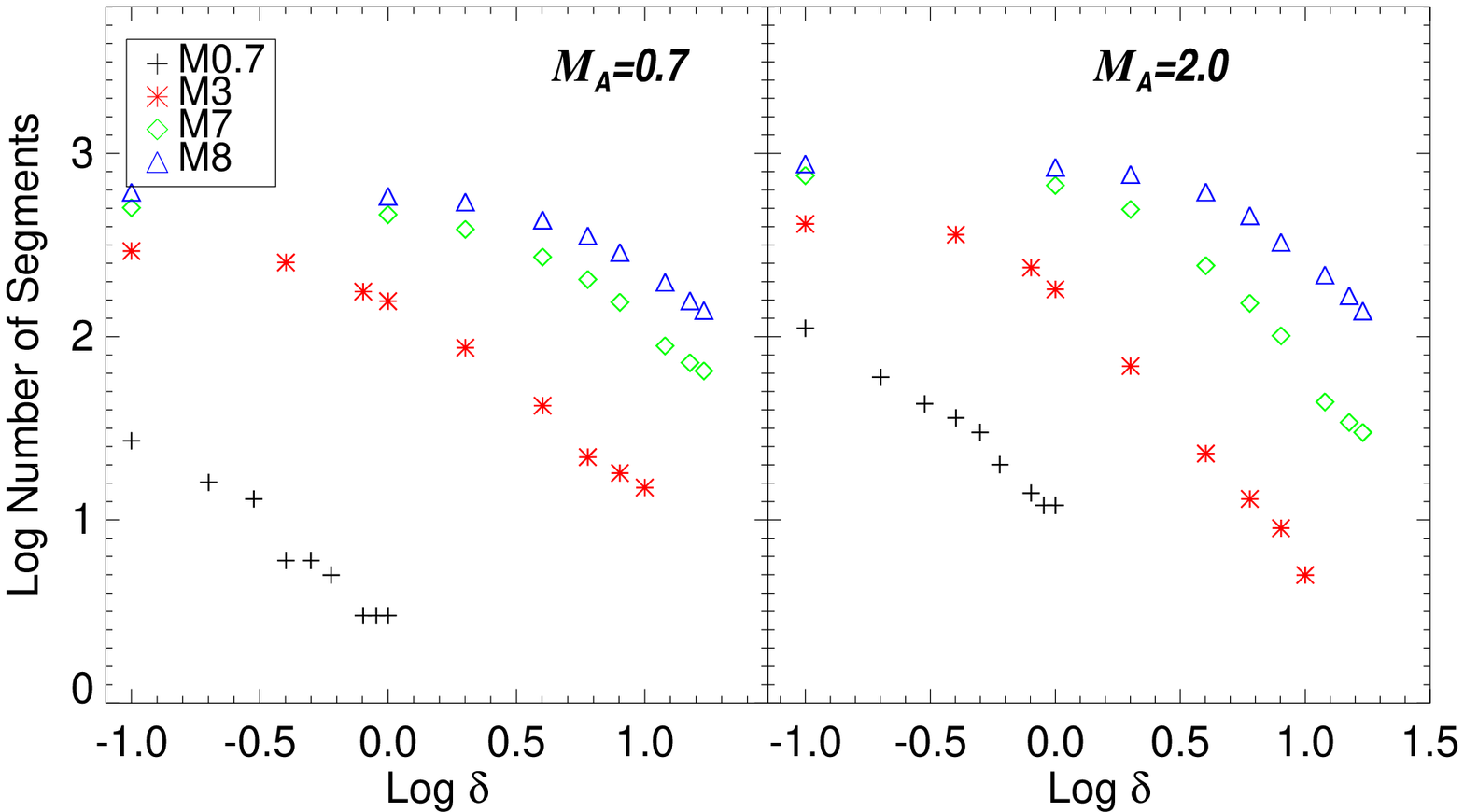}
\caption{Top: Total number of structures (leaves and branches) vs.~$\delta$.
Bottom: Number of segments from root to leaf on the largest branch of the tree vs. $\delta$.  Both plots are similar
to Figure \ref{fig:maxvsdelta}, only here we include the effects of thermal broadening. The color and symbol scheme used to represent different sonic Mach numbers is the same as that of Figure \ref{fig:maxvsdelta} for both top and bottom plots }
\label{fig:thermal}
\end{figure}

We repeat the analysis done in Figure \ref{fig:maxvsdelta}, using the models which now include thermal broadening, in Figure~\ref{fig:thermal}.  
We show the fitted slopes of the number of structures vs. $\delta$  from Figure \ref{fig:maxvsdelta} (top panel) as solid black lines accompanied by the numerical value of the slope. These black lines serve as a reference for which to compare with simulations without thermal broadening with 
the simulations that now include thermal broadening (colored symbols). 

As expected, the supersonic dendrograms are mostly unaffected by the inclusion of thermal broadening, since turbulent dominates the line profits.  We find the  number of structures vs. $\delta$ and corresponding slopes (top plot)  for the supersonic simulations to be very similar to those shown in Figure  \ref{fig:maxvsdelta}, with only a slight shallowing of the slopes.  Additionally the amount of hierarchical branching (bottom plot) is also similar.
However the subsonic PPV cube intensities are more drastically lowered as a result of the convolution with a broader Gaussian and thus the values of $\delta$ must be lowered as well. Additionally, the slopes seen for the subsonic simulations in the top plot of Figure~\ref{fig:thermal} are shallower then the slopes of the simulations with no thermal broadening included (comparison with solid lines referencing Figure \ref{fig:maxvsdelta}).  
Therefore when examining the the dendrograms of warm gas (such as HI and H$\alpha$) thermal broadening effects must also be taken into account.

\section{Discussion}
\label{disc}
Hierarchical tree diagrams are finding more applications in interstellar studies, not only to locate clumps and calculate their properties, but also for characterizing 
properties of the physics present in interstellar and molecular gas.  We used 
dendrograms to analyze how turbulence, magnetic fields and self-gravity shape the amount of structure and gas hierarchy in isothermal simulations. 
We examined the changes in the distribution
of the dendrogram as we vary the threshold parameter $\delta$. This is analogous to changing the corresponding threshold parameter in other techniques that rely on 
contouring thresholds, e.g. 
in the Genus analysis 
(see Chepurnov et al. 2009). By varying $\delta$ we obtained a new outlook on the technique; in particular, we found that the dendrogram distribution and hierarchy
have a strong dependency on the magnetization and compressibility of the gas and are sensitive to the amount of self-gravity.

\subsection{The Hierarchical Nature of MHD Turbulence}
The number of structures and the amount of hierarchy formed by MHD turbulence has interesting implications for the evolution of ISM clouds and for the star formation
problem.  In Section \ref{results} we found that more hierarchical structure and more overall structure was created in the presence of supersonic turbulence.
We also found that the inclusion of self-gravity enhanced these trends. 
The magnetic field also had a strong influence in the creation of hierarchical nesting in PPV space.
The relationship between the magnetization and the cloud dynamics is still not well understood, 
especially in regards to star formation.  Star forming clouds are known to be hierarchical in nature  and magnetized, but their exact Alfv\'enic nature is less clear.
The results from this work seem to suggest that very hierarchical clouds might tend towards being super-Alfv\'enic. 
Several authors have suggested a variety of  evidence for molecular clouds being super-Alfv\'enic.  This includes
the agreement of simulations and observations of Zeeman-splitting measurements, $B$ vs.$\rho$ relations, ${\cal M}_A$ vs. $\rho$ relations,
statistics of the extinction measurement etc. (Padoan \& Nordlund 1999; Lunttila et al. 2008; Burkhart et al. 2009; Crutcher et al. 2009; Collins et al. 2012).
Furthermore a study done by 
Burkhart et al. 2009 found that even in the presence of globally sub-Alfv\'enic turbulence, the highest density regions tend towards being locally super-Alfv\'enic.
This suggests that even in the case of globally sub-Alfv\'enic turbulence, the densest regions might be super-Alfv\'enic.
It is interesting that the dendrogram technique also points to super-Alfv\'enic turbulence as an avenue for hierarchical structure creation.
This provides motivation for the dendrogram technique to be applied to the observational data with varying threshold value $\delta$ in order
to see how the nature of the hierarchical structure and total number of structures change in the observations.

\subsection{Characterizing self-gravity and obtaining the Sonic and Alfv\'enic Mach Numbers from the Observations}
We provided a systematic study of the variations of the dendrogram thresholding parameter $\delta$
with the sonic and magnetic Mach numbers.  We also included a simulation with weak self-gravity (global virial number of 90) in order to investigate the influence gravity has on the observed hierarchy.   While real molecular clouds generally have virial numbers much lower than this value, we showed that the dendrogram is highly influenced by the inclusion of even weak self-gravity. The sonic and Alfv\'enic Mach numbers, as well as the virial parameter, are critical for understanding most processes in diffuse
and molecular gas, including the process of star formation. Thus, the dendrogram, with its sensitivity to these parameters,  provides a possible avenue of 
obtaining the characteristics of turbulence and the relative importance of self-gravity to turbulence in the ISM.  For example, the dendrogram, coupled with a virial analysis, was already used to compare the relative importance of self-gravity 
 within the L1148  GMC cloud using $^{13}$CO data in Rosolowsky et al. 2008 and  Goodman et al. 2009.

We view this work as a springboard for applying this technique to the observational data, which is why we addressed the issues of smoothing and thermal broadening in Section \ref{sec:app}.  It is clear that the relation between the dendrogram structures, their statistics and the thresholding value $\delta$, as explored in this work (for examples see Figures \ref{fig:maxvsdelta} and \ref{fig:sg5}), do not yield universal numbers, i.e.  they depend on the observational characteristics such as the velocity resolution and beams smoothing etc.  Therefore, in order to apply the dendrogram to the observations, we feel that it is necessary to define a fiducial dendrogram for the data where some information is known about turbulence, and in the case of self-gravitating clouds, the virial parameter.  The fiducial dendrogram can then be compared to other regions within the same data, which all contain the same observational constraints.  Similarly, for comparison of simulations and observations, the simulated observations must be tailored to the resolution of the observational data.

In order to obtain a fiducial region for further dendrogram analysis and to increase the reliability of the parameters found via the dendrogram,  it is advantageous to combine different techniques designed to investigate ISM turbulence.
For instance, by applying the VCA and VCS techniques to PPV data (see Lazarian 2009 for a review), one can obtain the velocity and 
density spectra of turbulence. While these measures are known to depend on ${\cal M}_s$ and to a lesser degree on ${\cal M}_A$ (see Beresnyak, 
Lazarian \& Cho 2005, Kowal, Lazarian \& Beresnyak 2007, Burkhart et al. 2009), the utility of the spectra is not limited to measuring these quantities. 
Spectra provide a unique way to investigate how the energy cascades between different scales, and show whether comparing observations with the simulations 
with a single scale of injection is reasonable.

The analysis of the anisotropies of correlations using velocity centroids provides an insight into media magnetization, i.e., it provides ${\cal M}_A$
(Lazarian et al. 2002, Esquivel \& Lazarian 2005), which is complementary to the dendrogram technique. Studies of the variance, skewness
and kurtosis of the PDFs (see Kowal, Lazarian \& Beresnyak 2007; Burkhart et al. 2009, 2010, 2012) provides measures of the sonic Mach number ${\cal M}_s$. 
Similarly, Tsallis statistics measures (Esquivel \& Lazarian 2010, Tofflemire et al. 2011) provide additional ways of estimating both ${\cal M}_s$  and ${\cal M}_A$.
We feel the approach to obtaining these parameters should be conducted with synergetic use of multiple tools, as was done in 
Burkhart et al. 2010 on the SMC.  The dendrogram is a unique tool as it can classify the hierarchical nature of the data and
 that it should be added to a standard set of statistical-tools for studies of ISM data.   

All these techniques provide independent ways of evaluating parameters of turbulence and therefore their application to the same data set provides
a  more reliable estimate of key parameters such as compressibility, magnetization, and degree of self-gravity.  
Dendrograms have some  advantages over other statistics designed to search for turbulence parameters, in that one can 
analyze the resulting tree diagram in many different ways, as highlighted in here and in previous works. 
 These include finding local maxima, calculating physical properties of dominate emission, exploring how those clumps
are connected in PPV, varying the threshold and calculating moments and level of hierarchy.    
Of course, one should keep in mind that the medium that we investigate observationally is far from simple. Multiple energy injection sources, for example, are not 
excluded. Thus, obtaining a similar answer with different techniques should provide us with additional confidence in our results. 

Finally, we should stress that for studies of astrophysical objects, the dendrogram and other statistical measures can be applied locally to different parts of the media.
 For instance,  Burkhart et al. (2010) did not characterize the entire SMC with one sonic Mach number. 
 Instead, several measures were applied  to parts of the SMC in order to obtain a distribution of the turbulence in the galaxy. A similar local scale selection
was applied also to the SMC in Chepurnov et al. (2008) using the Genus technique. 
Correlating the variations of the turbulence properties with observed properties of the media, e.g. star formation rate, should provide insight into how turbulence regulates many key astrophysical processes.

\section{Summary}
\label{con}
We investigated dendrograms of isothermal MHD simulations with varying levels of gravity, compressibility and magnetization using multiple values of the threshold parameter $\delta$. 
The dendrogram is a promising tool for studying both gas connectivity in the ISM as well as characterizing turbulence. In particular:
\begin{itemize}
\item We propose using statistical descriptions of dendrograms as a means to quantify the degree of hierarchy present in a PPV data cube.
\item Shocks, self-gravity, and super-Alfv\'enic turbulence create the most hierarchical structure in PPV space.  
\item The number of dendrogram structures depends primarily on the sonic number and self-gravity and secondarily on the global magnetization.
\item  The first four statistical moments of the distribution of dendrogram leaves and nodes have monotonic dependencies on the inclusion of self-gravity and
 the sonic and Alfv\'en Mach numbers over a range of $\delta$.
\item The dendrogram provides a convenient way of comparing  PPP to PPV in simulations.  Density structures are 
dominant in supersonic PPV and not in subsonic. Thus, it is more justifiable
to compare PPV to PPP when the gas is known to be supersonic.  
\end{itemize}

\acknowledgments
Authors thank Professor Diego Falceta-Gon\c{c}alves for the use of the self-gravitating simulation and useful discussions.
B.B also thanks Professor Jungyeon Cho for helpful discussion.
B.B. acknowledges support from the
NSF Graduate Research Fellowship and the NASA Wisconsin Space Grant
Institution. B.B. is thankful for valuable discussions and the use of the Dendrogui code via Chris Beaumont.
A.L. thanks  NSF grant AST 1212096 and both A.L. and B.B. thank the Center for Magnetic Self-Organization in Astrophysical and Laboratory Plasmas for financial support.
This work was completed  during the stay of A.L. as
Alexander-von-Humboldt-Preistr\"ager at the Ruhr-University Bochum.
A.G. acknowledges support from NSF Grant No. AST-0908159.
E.R. is supported by a Discovery Grant from NSERC of Canada.

\end{document}